\documentclass[12pt]{article}
\usepackage{amsmath}
\usepackage{myart}

\renewcommand{\theequation}{\arabic{section}.\arabic{equation}}
\normalbaselines
\input tcilatex.tex
\begin{document}

\author{Yuri A. Rylov}
\title{Dynamical methods of investigation in application to the Dirac
particle}
\date{Institute for Problems in Mechanics, Russian Academy of Sciences \\
101-1 ,Vernadskii Ave., Moscow, 119526, Russia \\
email: rylov@ipmnet.ru\\
Web site: {$http://rsfq1.physics.sunysb.edu/\symbol{126}rylov/yrylov.htm$}\\
or mirror Web site: {$http://195.208.200.111/\symbol{126}rylov/yrylov.htm$}}
\maketitle

\begin{abstract}
The Dirac particle $\mathcal{S}_{\mathrm{D}}$ is investigated by means of
dynamic methods, i.e. without a use of the principles of quantum mechanics.
It is shown that the Pauli particle $\mathcal{S}_{\mathrm{P}}$ and the
nonrelativistic approximation $\mathcal{S}_{\mathrm{nD}}$ of the Dirac
particle $\mathcal{S}_{\mathrm{D}}$ are different dynamic systems. $\mathcal{%
S}_{\mathrm{nD}}$ contains the high frequency degrees of freedom, which are
absent in the dynamic system $\mathcal{S}_{\mathrm{P}}$. It means that the
nonrelativistic Dirac particle $\mathcal{S}_{\mathrm{nD}}$ is composite
(i.e. it has internal degrees of freedom), whereas the Pauli particle $%
\mathcal{S}_{\mathrm{P}}$ is a pointlike particle with the spin. In the
absence of the electromagnetic field the world line of the classical Pauli
particle $\mathcal{S}_{\mathrm{Pcl}}$ is a timelike straight, whereas that
of the classical nonrelativistic Dirac particle $\mathcal{S}_{\mathrm{nDcl}}$
is a helix. The characteristic frequency $\Omega =2mc^{2}/\hbar $ of this
helix is the threshold frequency of the pair production. Using dynamic
methods, one shows freely that the Copenhagen interpretation, when the wave
function is a specific quantum object describing the state of individual
particle, is incompatible with the quantum mechanics formalism. Besides, it
is shown that the momentum distribution in quantum mechanics is in reality
the mean momentum distribution. Effectiveness of different investigation
strategies is discussed and compared.
\end{abstract}

\section{Introduction}

It is common practice to think, that the Pauli particle, i.e. the dynamic
system $\mathcal{S}_{\mathrm{P}}$, described by the Pauli equation, is a
nonrelativistic approximation of the Dirac particle, i.e. the dynamic system 
$\mathcal{S}_{\mathrm{D}}$, described by the Dirac equation \cite{D58}. In
reality, it appears that the nonrelativistic approximation $\mathcal{S}_{%
\mathrm{nD}}$ of the Dirac particle $\mathcal{S}_{\mathrm{D}}$ is a
composite particle, which is more complicated than the Pauli particle $%
\mathcal{S}_{\mathrm{P}}$, because it has additional degrees of freedom,
which absent at the Pauli particle. It can be explained as follows.

The Dirac equation is the system of four first order complex equations for
four complex dependent variables, whereas the Pauli equation is the system
of two complex first order equations for two complex dependent variables
(the order of differential equation is determined by the highest order of
time derivative). If the Pauli equation is a nonrelativistic approximation
of the Dirac equation, then why its order is lower. The reduction of the
order of differential equations is explained usually by the fact that the
coefficients before the highest temporal derivatives are small in the
nonrelativistic approximation. These terms are neglected and the order of
the differential equation is reduced. On the other hand, it is well known
that neglecting the highest derivatives, we loss high frequency solutions.
Indeed, if the temporal frequency is large enough, the term with the highest
derivative may become very large, even if the coefficient before the
derivative is small. It means that we may not neglect the highest
derivatives without producing a proper investigation.

In the present paper we produce such an investigation, which shows that one
cannot neglect the terms, which are connected with the internal structure of
the Dirac particle. Recently \cite{R2004} it has been shown that the
classical Dirac particle $\mathcal{S}_{\mathrm{Dcl}}$ has internal structure
(additional degrees of freedom). In the present paper it is shown that in
the nonrelativistic approximation the internal structure of the Dirac
particle takes place also.

From formal viewpoint the neglect of the high frequency solution is a
mathematical mistake in the investigation of the Dirac dynamic system $%
\mathcal{S}_{\mathrm{D}}$, and a very interesting question arises. Why was
the mathematical mistake in the transition to the nonrelativistic
approximation remaining to be unnoticed for eighty years after invention of
the Dirac equation? The answer is very simple. Nobody looked for this
mistake. The quantum theory developed by means of the experimental-fitting
methods, when the logical structure of a theory was a secondary
circumstance. It was necessary to explain the enigmatic microcosm by any
means. As a result the quantum theory is founded on the enigmatic quantum
principles, which are nonrelativistic. There is nothing bad in application
of the experimental-fitting method for explanation of concrete experiments
and experimental data, because this method admits one to introduce new
concepts, which are characteristic for the considered physical phenomena.
However, it leads to undesirable consequences, when the experimental-fitting
method is applied to a construction of a physical theory. In this case the
method turns into a theoretical-fitting method. Application of the
theoretical-fitting method to a construction of a theory is ineffective,
because it admits one to introduce new hypotheses and new concepts, but it
does not admit one to establish logical connections between different
concepts and explain some concepts via other more fundamental concepts. The
main goal of a physical theory is a determination of logical connections
between the physical concepts and deduction of all concepts via some
fundamental concepts. Any progress in reduction of the number of the
fundamental concepts and explanation of other concepts in terms of the
fundamental concepts is a real progress of a physical theory. Introducing
new hypotheses, the experimental-fitting method enables to introduce new
concepts, but it does not admit one to establish logical connection between
the physical concepts and reduce the number of fundamental basic concepts.
One should not use the theoretical-fitting method to determine the logical
connections between the physical concepts and to construct a satisfactory
theory . Instead of the fitting method one should use the Newtonian
deductive method with its slogan "Hypotheses non fingo".

If we have serious problems with new physical phenomena, what are we to do?
According to the Newtonian slogan we should look for mistakes in the
existing physical theory, find them and correct. The finding of mistakes in
the foundation of the existing physical theory is a very difficult problem,
because these mistakes appear to be located in other branches of science
(geometry, theory of dynamic systems and theory of stochastic systems). In
the beginning of the 20th century, when the quantum theory arose, these
mistakes were not discovered. The researchers of the 20th century were
forced to invent additional hypotheses (quantum principles), which could
compensate unknown mistakes in the foundation of the physical theory of
microcosm phenomena. In the same way Ptolemeus constructed his doctrine of
the celestial mechanics, where the mistake concerning the Earth motion was
compensated by additional suppositions. The Ptolemaic doctrine described
correctly the motion of planets, but it cannot be used for discovery of the
Newtons gravitation law and for calculation of trajectories of rockets in
their travel to other planets. In the same way the contemporary quantum
theory describes correctly the atomic spectra and other nonrelativistic
quantum phenomena, but it fails in the description of the specific
relativistic quantum phenomena. The Ptolemaic doctrine (as well as the
contemporary quantum theory) was a list of the prescriptions, which lead to
true experimentally tested predictions. But not all these prescription were
connected logically between themselves. Some of them were compatible only in
some region of parameters of the theory, and the doctrine cannot be applied
outside this region. List of these prescriptions did not form a logical
structure, and the reason was an incorrect fundamental supposition,
concerning the Earth motion. The list of the Ptolemaic prescriptions did not
form a logical structure. It could be applied only to the planet motion, but
it could not be extended to motion of other celestial bodies (comets,
rockets). The Copernicus doctrine was a logical structure, because it did
not contain mistake concerning the Earth motion. It could be extended to the
motion of any celestial bodies.

Analogous situation takes place in the contemporary quantum theory, which
also forms a list of prescriptions, but not a logical structure. These
prescriptions work very well in the nonrelativistic phenomena of microcosm,
but one fails to extend them to relativistic phenomena of microcosm. The
reason of this failure is conditioned by the incorrect statements (mistakes)
in the foundation of the contemporary quantum theory.

We list the incorrect statements (mistakes), which must be corrected for a
construction of a satisfactory theory of microcosm phenomena.

\begin{enumerate}
\item \textit{The straight is a one-dimensional line in any space-time
geometry}. This statement forbids space-time geometries, where the motion of
free particles is primordially stochastic.

\item \textit{Any statistical description is produced in terms of the
probability theory.} This statement forbids the dynamical conception of the
statistical description, which does not use the concept of the probability
density as a main concept of the statistical description.

\item \textit{The free particle Hamiltonian function }$H$\textit{\ and its
energy }$E$\textit{, taken with the opposite sign, always coincide }$\left(
E=-H\right) $.
\end{enumerate}

The fourth problem, which should be overcome, is not a mistake. It was a
purely mathematical problem. Without solving this problem, one cannot obtain
the correct interpretation of the wave function as a method of an ideal
fluid description. We consider this problem in the second section.

Two first points concern the quantum theory as a whole, whereas the third
point concerns only relativistic quantum theory. It concerns the problem of
the pair production and depreciates many papers on the relativistic quantum
field theory. We show this in the example of the second quantization of the
nonlinear Klein-Gordon equation 
\begin{equation}
\partial _{0}^{2}\psi -\mathbf{\nabla }^{2}\psi +m^{2}\psi =:\lambda \psi
^{+}\psi \psi :  \label{a0.1}
\end{equation}%
where the speed of the light $c=1$, the quantum constant $\hbar =1$, and $%
\lambda $ is the constant of self-action.

At the secondary quantization the nonlinear term in rhs of (\ref{a0.1})
provides the pair production, if one imposes some additional constraint \cite%
{GJ68,GJ70,GJ970,GJ72}

\begin{equation}
\psi P_{k}-P_{k}\psi =-i\hbar \frac{\partial \psi }{\partial x^{k}},\qquad
P^{k}=\int T^{0k}d\mathbf{x,\qquad }k=0,1,2,3  \label{a0.2}
\end{equation}%
where $T^{ik}$ is the energy-momentum tensor. Conventionally the condition (%
\ref{a0.2}) is considered to be the condition, which is necessary for the
secondary quantization. Nobody does not consider the conditions (\ref{a0.2})
as some additional constraints, which are not necessary for the secondary
quantization, and nobody tests compatibility of constraints (\ref{a0.2})
with the dynamic equation (\ref{a0.1}). However, the secondary quantization
of the equation (\ref{a0.1}) is possible without imposition of constraints (%
\ref{a0.2}) \cite{R2001a}. It means that the conditions (\ref{a0.2}) are
additional constraints and compatibility of constraints (\ref{a0.2}) with
the dynamic equation (\ref{a0.1}) is to be tested. The test has been made in 
\cite{R2001a}. It has been shown, that the relations (\ref{a0.2}) and (\ref%
{a0.1}) are compatible only in the case, when the self-action constant $%
\lambda =0$, and the dynamic equation (\ref{a0.1}) is linear.

Thus, although the statement of the pair production problem in the form of
two relations (\ref{a0.1}) and (\ref{a0.2}) leads to the pair production
effect \cite{GJ68,GJ70,GJ970,GJ72}, but this result is not reliable, because
the statement of the problem is inconsistent. Besides, the mathematical
formalism is imperfect, because it uses perturbation theory and
renormalization. Combination of the inconsistent statement of the problem
with the imperfect mathematical technique admits one to obtain any desirable
result (in the given case the effect of pair production).

On the other hand, the secondary quantization of the equation (\ref{a0.1})
without imposition of (\ref{a0.2}) provides the consistent statement of the
problem with the perfect nonperturbative mathematical technique (without
renormalization). However, the pair production effect is absent at such a
consistent statement of the problem \cite{R2001a}.

What is the physical ground of the constraint (\ref{a0.2}), which leads to
the pair production effect? If $k=0$ the relation (\ref{a0.2}) describes the
well known fact that for the free particle $H=-E$, where $H$ is the Hamilton
function, defined as the quantity canonically conjugate to the time $t$, and 
$E$ is the particle energy, defined as an integral of the component $T^{00}$
of the energy-momentum tensor. But this relation is valid only in the case,
when the pair production is absent \cite{R70}. In the general case, when
there is the pair production, the imposition of constraint (\ref{a0.2})
means that the description is produced in terms of particles and
antiparticles, which are considered as different dynamic systems \cite{R70}.
The number of objects is indefinite, and one is forced to use the
perturbative methods. On the contrary, absence of the constraint (\ref{a0.2}%
) means that the description is produced in terms of world lines, which are
considered as the fundamental objects of dynamics. The number of these
extended objects is fixed, and one may use nonperturbative methods of
investigation (see details in \cite{R2001a}).

Dynamics, where the dynamic system (particle) exists only some time and
disappears at some time moment after collision with the dynamic anti-system
(antiparticle), is inconsistent. The technique of classical dynamic systems
does not admit one to use such a dynamical description. However, the same
tecnique admits one to describe this collision, if a particle and an
antiparticle are different states of the same physical object (world line),
and evolution of the dynamic system is determined by a parameter changing
monotone along the world line. In this case the collision leads only to a
transition from one state to another. Mathematical technique of quantum
theory also cannot overcome the difficulty, connected with particle and
antiparticle as different dynamic systems. The belief, that we can overcome
this difficulty, introducing creation and annihilation operators, is
delusive. It is a reason, why the relations (\ref{a0.1}) and (\ref{a0.2})
are incompatible, and it is the third point in the list of mistakes.

It means that a simple addition of the nonlinear term to the linear
Klein-Gordon equation does not provide the pair production effects. The
reasons, generating the pair production, have a more complicated structure,
than a simple product of the creation and annihilation operators. Besides,
these reasons have a classical analog in the form of specific force fields.
(See for details \cite{R2003}).

We see in the considered example, that the quantum principles do not work in
application to the relativistic quantum systems, or at least, they are not
effective in application to them. There is a hope that the relativistic
quantum systems can be investigated more effectively by dynamical methods,
which do not use the quantum principles.

The above-mentioned mistakes look very simple, and it is very difficult to
believe that a correction of these incorrect statements could lead to a
construction of a satisfactory theory of the microcosm phenomena. However,
these mistakes underlie of the contemporary quantum theory of microcosm
phenomena, and correction of them is very important for further development
of the microcosm phenomena theory. In particular, correction of the first
mistake leads to construction of the new conception of geometry \cite{R2005}
and to a revision of our space-time conception in the microcosm. The scale
of this revision is comparable with the scale of revision connected with
appearance of the relativity theory \cite{R2003a}. Correction of the second
mistake lead to a construction of the dynamical conception of the
statistical description. Dynamical methods of this conception are used in
the present paper. But we apply the dynamical methods without a reference to
their physical foundation. The fact, that the dynamical methods have
appeared in accordance with the Newtonian investigation strategy as a result
of correction of a mistake, is very important from the logical viewpoint.

Unfortunately, this fact is of no importance for contemporary pragmatic
theorists, educated on the experimental-fitting method of investigation.
They do not believe in any foundation and trust only in effectiveness of the
applied investigation methods. In the given case I prefer to use their
rules, in order the paper were transparent for most readers, educated on the
experimental-fitting method of investigation.

As concerns the third mistake, it is not yet corrected properly in the sense
that the effective theory of the pair production effect is not yet
constructed. It is clear only, that application of the quantum principles in
solution of this problem leads to the blind alley.

In this paper we show that the dynamical methods of investigation (without a
use of quantum principles) are founded logically. Besides, they are more
effective in application to the investigation of the Dirac particle, than
the conventional methods, based on the application of quantum principles.

\section{Dynamical methods of investigation}

We use a more developed mathematical technique for a description of quantum
systems. This technique supposes that all essential information on the
quantum dynamical system is contained in the dynamic system itself. Such
specific quantum concepts as the wave function and principles of quantum
mechanics appear to be only the means of description. The wave function as
the means of description may be applied to both quantum and classical
dynamic systems. But the quantum principles may be applied only to quantum
dynamic systems, because they contains some constraints, which are not
satisfied for classical systems. The quantum system and classical system
distinguish dynamically (in additional terms in the action), but not in the
way of description. This fact becomes to be clear, when both systems are
described in the same terms. For instance, the quantum system and the
corresponding classical system may be described in terms of the wave
function, or both systems may be described in terms of the particle position
and momentum. The difference between the various methods of description lies
only in the convenience of their application. A use of the wave function is
effective in description of quantum systems, because in this case the
dynamic equations are linear. On the contrary, the description in terms of
the particle position is convenient for description of classical systems,
where dynamic equations are ordinary differential equations.

Progress in the development of the mathematical technique has a \textit{%
mathematical ground: integration of dynamic equations}. This pure
mathematical achievement has physical consequences. It appears that the
quantum mechanics may be considered to be a statistical description of
randomly moving particles. We underline that we investigate well known
quantum systems, and all new results are \textit{corollaries of the more
developed methods of investigation}. It is meaningless to argue against the
new obtained results by a reference to experimental data, because such
arguments are arguments against the considered dynamic systems, but not
against the methods of investigation. Experimental data may not be arguments
against the mathematical methods of investigation in principle. As to the
investigated dynamic systems, we admit that they may be imperfect and need
an improvement, but this problem lies outside the framework of the paper.

We show new mathematical methods of investigation in the simple example of
the Schr\"{o}dinger particle $\mathcal{S}_{\mathrm{S}}$, i.e. the dynamic
system $\mathcal{S}_{\mathrm{S}}$, described by the Schr\"{o}dinger equation.

The action for the free nonrelativistic quantum particle $\mathcal{S}_{%
\mathrm{S}}$ has the following form 
\begin{equation}
\mathcal{S}_{\mathrm{S}}:\qquad \mathcal{A}_{\mathrm{S}}\left[ \psi ,\psi
^{\ast }\right] =\int \left\{ \frac{i\hbar }{2}\left( \psi ^{\ast }\partial
_{0}\psi -\partial _{0}\psi ^{\ast }\cdot \psi \right) -\frac{\hbar ^{2}}{2m}%
\mathbf{\nabla }\psi ^{\ast }\mathbf{\nabla }\psi \right\} dtd\mathbf{x}
\label{a1.2}
\end{equation}%
where $\psi =\psi \left( t,\mathbf{x}\right) $ is a complex one-component
wave function, $\psi ^{\ast }=\psi ^{\ast }\left( t,\mathbf{x}\right) $ is
the complex conjugate to $\psi $, and $m$ is the particle mass. The action (%
\ref{a1.2}) generates dynamic equations 
\begin{equation}
i\hbar \partial _{0}\psi =-\frac{\hbar ^{2}}{2m}\mathbf{\nabla }^{2}\psi
,\qquad -i\hbar \partial _{0}\psi ^{\ast }=-\frac{\hbar ^{2}}{2m}\mathbf{%
\nabla }^{2}\psi ^{\ast }  \label{a1.2a}
\end{equation}%
The 4-current $j^{k}$ and the energy-momentum tensor $T_{l}^{k}$ are the
canonical quantities associated with the action $\mathcal{A}_{\mathrm{S}}%
\left[ \psi ,\psi ^{\ast }\right] $. They are determined by the relations 
\begin{equation}
j^{k}=\left\{ \rho ,\mathbf{j}\right\} =\frac{i}{\hbar }\left( \frac{%
\partial \mathcal{L}}{\partial \left( \partial _{k}\psi ^{\ast }\right) }%
\psi ^{\ast }-\frac{\partial \mathcal{L}}{\partial \left( \partial _{k}\psi
\right) }\psi \right) =\left\{ \psi ^{\ast }\psi ,-\frac{i\hbar }{2m}\left(
\psi ^{\ast }\mathbf{\nabla }\psi -\mathbf{\nabla }\psi ^{\ast }\cdot \psi
\right) \right\}  \label{a1.3}
\end{equation}%
\begin{equation}
T_{l}^{k}=\frac{\partial \mathcal{L}}{\partial \left( \partial _{k}\psi
^{\ast }\right) }\partial _{l}\psi ^{\ast }+\frac{\partial \mathcal{L}}{%
\partial \left( \partial _{k}\psi \right) }\partial _{l}\psi -\delta _{l}^{k}%
\mathcal{L},\qquad k,l=0,1,2,3  \label{a1.4}
\end{equation}%
where $\mathcal{L}$ is the Lagrangian density for the action (\ref{a1.2}) 
\begin{equation}
\mathcal{L}=\frac{i\hbar }{2}\left( \psi ^{\ast }\partial _{0}\psi -\partial
_{0}\psi ^{\ast }\cdot \psi \right) -\frac{\hbar ^{2}}{2m}\mathbf{\nabla }%
\psi ^{\ast }\mathbf{\nabla }\psi  \label{a1.4a}
\end{equation}

The dynamic system $\mathcal{S}_{\mathrm{S}}$ is determined completely by
dynamic equations (\ref{a1.2a}) and expressions (\ref{a1.3}), (\ref{a1.4})
for the 4-current and the energy-momentum tensor. Only connection between
the particle and the wave functions is not described by these relations.
This connection is described by means of the relations

\begin{equation}
\left\langle F\left( \mathbf{x},\mathbf{p}\right) \right\rangle =B\int \func{%
Re}\left\{ \psi ^{\ast }F\left( \mathbf{x},\mathbf{\hat{p}}\right) \psi
\right\} d\mathbf{x,\qquad \hat{p}}=\mathbf{-i\hbar \mathbf{\nabla },\qquad }%
B=\left( \int \psi ^{\ast }\psi d\mathbf{x}\right) ^{-1}  \label{b1.5}
\end{equation}
which define the mean value $\left\langle F\left( \mathbf{x},\mathbf{p}%
\right) \right\rangle $ of any function $F\left( \mathbf{x},\mathbf{p}%
\right) $ of the particle coordinates $\mathbf{x}$ and momentum $\mathbf{p}$%
. Application of the rules (\ref{b1.5}) is restricted by some conditions.
They demand that the dynamic equations be linear and the wave function be a
vector in the Hilbert space of states. We shall refer to the relations (\ref%
{b1.5}) together with the restrictions imposed on its applications as the
quantum principles, because von Neumann has shown \cite{N32}, that the
quantum mechanics can be deduced from relations of the type (\ref{b1.5}),
provided they are valid for all observable quantities. Thus, the
interpretation of the wave function is carried out on the basis of the
quantum principles, which are something external with respect to the dynamic
system $\mathcal{S}_{\mathrm{S}}$.

In reality, the quantum principles are not necessary for interpretation of
the dynamic system $\mathcal{S}_{\mathrm{S}}$. It is sufficient to make a
proper change of dynamic variables and to describe the dynamic system $%
\mathcal{S}_{\mathrm{S}}$ in terms of the particle coordinates $\mathbf{x}$.
Such a description does not contain the enigmatic wave function, whose
meaning is unclear, and one does not need the quantum principles (\ref{b1.5}%
) for its interpretation. The Schr\"{o}dinger particle $\mathcal{S}_{\mathrm{%
S}}$ is a partial case of the generalized Schr\"{o}dinger particle $\mathcal{%
S}_{\mathrm{gS}}$, which is the dynamic system $\mathcal{S}_{\mathrm{gS}}$,
described by the action 
\begin{equation}
\mathcal{A}_{\mathrm{gS}}[\psi ,\psi ^{\ast }]=\int \left\{ \frac{i\hbar }{2}%
\left( \psi ^{\ast }\partial _{0}\psi -\partial _{0}\psi ^{\ast }\cdot \psi
\right) -\frac{\hbar ^{2}}{2m}\mathbf{\nabla }\psi ^{\ast }\mathbf{\nabla }%
\psi +\frac{\hbar ^{2}}{8m}\sum\limits_{\alpha =1}^{\alpha =3}(\mathbf{%
\nabla }s_{\alpha })^{2}\rho \right\} \mathrm{d}^{4}x  \label{b1.5a}
\end{equation}
\begin{equation}
\rho \equiv \psi ^{\ast }\psi ,\qquad \mathbf{s}\equiv \frac{\psi ^{\ast }%
\mathbf{\sigma }\psi }{\rho },\qquad \mathbf{\sigma }=\{\sigma _{\alpha
}\},\qquad \alpha =1,2,3,  \label{b1.5b}
\end{equation}
Here $\psi =\left( _{\psi _{2}}^{\psi _{1}}\right) $, $\psi ^{\ast }=\left(
\psi _{1}^{\ast },\psi _{2}^{\ast }\right) $ is the two-component wave
function, and $\sigma _{\alpha }$ are the Pauli matrices. The 4-current is
defined by the relation (\ref{a1.3}) with two-component wave function $\psi $%
. In the case, when components $\psi _{1}$ and $\psi _{2}$ are linear
dependent (for instance, $\psi =\left( _{0}^{\psi _{1}}\right) $), the mean
spin vector $\mathbf{s}=$const, and the last term in the action (\ref{b1.5a}%
) vanishes. In this case the dynamic system $\mathcal{S}_{\mathrm{gS}}$
turns into the dynamic system (\ref{a1.2}).

One can show, that the dynamic system $\mathcal{S}_{\mathrm{gS}}$ is another
representation of the dynamic system $\mathcal{E}\left[ \mathcal{S}_{\mathrm{%
st}}\right] $, i.e. the action for $\mathcal{S}_{\mathrm{gS}}$ can be
obtained from the action for the dynamic system $\mathcal{E}\left[ \mathcal{S%
}_{\mathrm{st}}\right] $ by means of a proper change of variables \cite{R99}.

The dynamic system $\mathcal{E}\left[ \mathcal{S}_{\mathrm{st}}\right] $ is
a statistical ensemble of stochastic particles $\mathcal{S}_{\mathrm{st}}$.
It is described by the action 
\begin{equation}
\mathcal{E}\left[ \mathcal{S}_{\mathrm{st}}\right] :\qquad \mathcal{A}_{%
\mathcal{E}\left[ \mathcal{S}_{\mathrm{st}}\right] }\left[ \mathbf{x,u}_{%
\mathrm{st}}\right] =\int \left\{ \frac{m}{2}\left( \frac{d\mathbf{x}}{dt}%
\right) ^{2}+\frac{m}{2}\mathbf{u}_{\mathrm{st}}^{2}-\frac{\hbar }{2}\mathbf{%
\nabla u}_{\mathrm{st}}\right\} dtd\mathbf{\xi }  \label{a1.22}
\end{equation}
where $\mathbf{u}_{\mathrm{st}}=\mathbf{u}_{\mathrm{st}}\left( t,\mathbf{x}%
\right) $ is a vector function of arguments $t,\mathbf{x}$ (not of $t,%
\mathbf{\xi }$), and $\mathbf{x}=\mathbf{x}\left( t,\mathbf{\xi }\right) $
is a 3-vector function of independent variables $t,\mathbf{\xi =}\left\{ \xi
_{1,}\xi _{2},\xi _{3}\right\} $. Dynamic equations for the dynamic system $%
\mathcal{E}\left[ \mathcal{S}_{\mathrm{st}}\right] $ are obtained as a
result of variation of the action (\ref{a1.22}) with respect to dependent
dynamic variables $\mathbf{x,u}_{\mathrm{st}}$. In the action (\ref{a1.22})
the variables $\mathbf{\xi }$ label stochastic systems $\mathcal{S}_{\mathrm{%
st}}$, constituting the statistical ensemble. The operator $\mathbf{\nabla }$
is defined in the space of coordinates $\mathbf{x}$ by the relation 
\begin{equation}
\mathbf{\nabla =}\left\{ \partial _{1},\partial _{2},\partial _{3}\right\} 
\mathbf{\equiv }\left\{ \frac{\partial }{\partial x^{1}},\frac{\partial }{%
\partial x^{2}},\frac{\partial }{\partial x^{3}}\right\}  \label{a1.22a}
\end{equation}
The 3-vector $\mathbf{u}_{\mathrm{st}}$ describes the mean value of the
stochastic component of the particle motion, which is considered to be a
function of the variables $t,\mathbf{x}$. The first term $\frac{m}{2}\left( 
\frac{d\mathbf{x}}{dt}\right) ^{2}$ describes the energy of the regular
component of the stochastic particle motion. The second term $m\mathbf{u}_{%
\mathrm{st}}^{2}/2$ describes the energy of the random component of
velocity. The components $\frac{d\mathbf{x}}{dt}$ and $\mathbf{u}_{\mathrm{st%
}}$ of the total velocity are connected with different degrees of freedom,
and their energies should be added in the expression for the Lagrange
function density. The last term $-\hbar \mathbf{\nabla u}_{\mathrm{st}}/2$
describes interplay between the velocity $\frac{d\mathbf{x}}{dt}$ of the
regular component and the random one $\mathbf{u}_{\mathrm{st}}$.

The action (\ref{a1.22}) is a sum (integral) of actions for independent
stochastic systems $\mathcal{S}_{\mathrm{st}}$, labelled by the parameters $%
\mathbf{\xi }=\left\{ \xi _{1},\xi _{2},\xi _{3}\right\} $. Any stochastic
system $\mathcal{S}_{\mathrm{st}}$ is a stochastic particle, whose state is
described by its coordinate $\mathbf{x}\left( t\right) $. The action for the
stochastic system $\mathcal{S}_{\mathrm{st}}$ is obtained from the action (%
\ref{a1.22}) for $\mathcal{E}\left[ \mathcal{S}_{\mathrm{st}}\right] $. It
has the form 
\begin{equation}
\mathcal{S}_{\mathrm{st}}:\qquad \mathcal{A}_{\mathcal{S}_{\mathrm{st}}}%
\left[ \mathbf{x,u}_{\mathrm{st}}\right] =\int \left\{ \frac{m}{2}\left( 
\frac{d\mathbf{x}}{dt}\right) ^{2}+\frac{m}{2}\mathbf{u}_{\mathrm{st}}^{2}-%
\frac{\hbar }{2}\mathbf{\nabla u}_{\mathrm{st}}\right\} dt  \label{a1.23}
\end{equation}%
where $\mathbf{x}=\mathbf{x}\left( t\right) $. In reality, the action (\ref%
{a1.23}) is not well defined mathematically, if $\hbar \neq 0$. It is only
symbolic, because the operator (\ref{a1.22a}) is defined in the vicinity of
the point $\mathbf{x}$, but not at the point $\mathbf{x}$ itself. As a
result the dynamic equations for the stochastic system $\mathcal{S}_{\mathrm{%
st}}$ do not exist, if $\hbar \neq 0$. This fact agrees with the
stochasticity of $\mathcal{S}_{\mathrm{st}}$. By definition the system $%
\mathcal{S}_{\mathrm{st}}$ is stochastic, if there exist no dynamic euations
for $\mathcal{S}_{\mathrm{st}}$. If we cut off interaction with the
stochastic agent, setting $\hbar =0$ in the action (\ref{a1.23}) (or$\ $%
remove two last terms), we obtain the well defined action for the free
nonrelativistic deterministic particle $\mathcal{S}_{\mathrm{d}}$%
\begin{equation}
\mathcal{S}_{\mathrm{d}}:\qquad \mathcal{A}_{\mathcal{S}_{\mathrm{d}}}\left[ 
\mathbf{x,u}_{\mathrm{st}}\right] =\int \left\{ \frac{m}{2}\left( \frac{d%
\mathbf{x}}{dt}\right) ^{2}+\frac{m}{2}\mathbf{u}_{\mathrm{st}}^{2}\right\}
dt,\qquad \mathbf{x}=\mathbf{x}\left( t\right)  \label{a1.23a}
\end{equation}

The Schr\"{o}dinger particle $\mathcal{S}_{\mathrm{S}}$ (\ref{a1.2}) is a
partial case of the dynamic system $\mathcal{E}\left[ \mathcal{S}_{\mathrm{st%
}}\right] $ (\ref{a1.22}), whereas the generalized Schr\"{o}dinger particle $%
\mathcal{S}_{\mathrm{gS}}$ (\ref{b1.5a}) coincide with the dynamic system $%
\mathcal{E}\left[ \mathcal{S}_{\mathrm{st}}\right] $ (\ref{a1.22}). The
action (\ref{b1.5a}) may be obtained from the action (\ref{a1.22})
mathematically by means of a proper change of variables. (see Appendix A).

Interpretation of the dynamic system (\ref{a1.22}) is very simple, but
dynamic equations for $\mathcal{E}\left[ \mathcal{S}_{\mathrm{st}}\right] $
are rather complicated. They have the form 
\begin{equation}
\frac{\delta \mathcal{A}_{\mathcal{E}\left[ \mathcal{S}_{\mathrm{st}}\right]
}}{\delta \mathbf{x}}=-m\frac{d^{2}\mathbf{x}}{dt^{2}}+\mathbf{\nabla }%
\left( \frac{m}{2}\mathbf{u}_{\mathrm{st}}^{2}-\frac{\hbar }{2}\mathbf{%
\nabla u}_{\mathrm{st}}\right) =0  \label{b1.4}
\end{equation}%
\begin{equation}
\frac{\delta \mathcal{A}_{\mathcal{E}\left[ \mathcal{S}_{\mathrm{st}}\right]
}}{\delta \mathbf{u}_{\mathrm{st}}}=m\mathbf{u}_{\mathrm{st}}\rho +\frac{%
\hbar }{2}\mathbf{\nabla }\rho =0,  \label{b1.7}
\end{equation}%
where $\rho $ is the function of derivatives of $\mathbf{x}$ with respect to 
$\mathbf{\xi }=\left\{ \xi _{1},\xi _{2},\xi _{3}\right\} $, determined by
the relation 
\begin{equation}
\rho =\left[ \frac{\partial \left( x^{1},x^{2},x^{3}\right) }{\partial
\left( \xi _{1},\xi _{2},\xi _{3}\right) }\right] ^{-1}=\frac{\partial
\left( \xi _{1},\xi _{2},\xi _{3}\right) }{\partial \left(
x^{1},x^{2},x^{3}\right) }  \label{b1.9}
\end{equation}%
Resolving the relation (\ref{b1.7}) with respect to $\mathbf{u}_{\mathrm{st}%
} $ in the form 
\begin{equation}
\mathbf{u}_{\mathrm{st}}=-\frac{\hbar }{2m}\mathbf{\nabla }\ln \rho ,
\label{b1.10}
\end{equation}%
and eliminating $\mathbf{u}_{\mathrm{st}}$ from (\ref{b1.4}), we obtain 
\begin{equation}
m\frac{d^{2}\mathbf{x}}{dt^{2}}=-\mathbf{\nabla }U\left( \rho ,\mathbf{%
\nabla }\rho \right) ,\qquad U\left( \rho ,\mathbf{\nabla }\rho \right) =%
\frac{\hbar ^{2}}{8m}\left( \frac{\left( \mathbf{\nabla }\rho \right) ^{2}}{%
\rho ^{2}}-2\frac{\mathbf{\nabla }^{2}\rho }{\rho }\right)  \label{b1.11}
\end{equation}%
Thus, dynamic equations, generated by the action (\ref{a1.22}), describe the
regular motion component of any particle $\mathcal{S}_{\mathrm{st}}$, as a
motion in a very complicated potential field $U$, depending on the
distribution of all particles of the statistical ensemble $\mathcal{E}\left[ 
\mathcal{S}_{\mathrm{st}}\right] $. Of course, the trajectories $\mathbf{x}=%
\mathbf{x}\left( t,\mathbf{\xi }\right) $ do not describe the motion of
individual stochastic particles. They describe only statistical average
motion of stochastic particles. The situation reminds situation in the gas
dynamics. The dynamic equations of the gas dynamics describe the motion of
the "gas particles", which contain many molecules. Motion of the gas
molecules is random and chaotic. It cannot be described by the gas dynamics
equations, which describe only regular component of the molecule motion.

Note, that the term $\frac{m}{2}\mathbf{u}_{\mathrm{st}}^{2}$ in (\ref{a1.23}%
) looks as a kinetic energy, but according to (\ref{b1.10}) it does not
depend on the temporary derivative $\mathbf{\dot{x}}$, and in dynamic
equations it acts as a potential energy.

The statistical ensemble (\ref{a1.22}) may be considered to be some fluid.
We may speak about the flow of the statistical ensemble $\mathcal{E}\left[ 
\mathcal{S}_{\mathrm{st}}\right] $, keeping in mind, that dynamic equation (%
\ref{b1.11}) for the dynamic system $\mathcal{E}\left[ \mathcal{S}_{\mathrm{%
st}}\right] $ may be interpreted as hydrodynamic equation for some "quantum"
fluid.

On the contrary, the dynamic equations, generated by the action (\ref{a1.2}%
), are linear and rather simple, whereas their interpretation is very
complicated, because it uses the principles of quantum mechanics (\ref{b1.5}%
). Thus, the description by means of the action (\ref{a1.22}) admits a
simple interpretation, but dynamic equations are very complicated for a
solution.

If the actions (\ref{a1.2}) and (\ref{a1.22}) describe the same dynamic
system (further for brevity we shall speak about equivalence of dynamic
systems (\ref{a1.2}) and (\ref{a1.22}), although in reality the action (\ref%
{a1.2}) is only a partial case of the action (\ref{a1.22})), it is
reasonable to use the dynamic system $\mathcal{E}\left[ \mathcal{S}_{\mathrm{%
st}}\right] $ as starting point for the statement of the problem and for
interpretation of the results obtained, whereas the dynamic system $\mathcal{%
S}_{\mathrm{S}}$ will be used only for solution of dynamic equations, which
have a simple form in terms of the wave function.

Why was this evident circumstance not used before? Why was the problem of
the stochastic motion of microparticles stated in terms of enigmatic wave
function? The answer is very simple. The connection between two different
forms (\ref{a1.2}) and (\ref{a1.22}) of the action for the Schr\"{o}dinger
particle \textit{has not been known for a long time. }

It is known, that the Schr\"{o}dinger equation can be written in the
hydrodynamical form \cite{M26}. D. Bohm \cite{B52} used this circumstance
for the hydrodynamic interpretation of quantum mechanics. But it was only
interpretation of the quantum principles in the hydrodynamical terms. He
failed to eliminate the quantum principles and the wave function from the
foundation of the quantum mechanics, and the wave function remained to be an
enigmatic object -- the vector in the Hilbert space. One failed to connect
the wave function with the hydrodynamic variables: the density $\rho $ and
the velocity $\mathbf{v}$. In more exact terms the connection between the
wave function and hydrodynamic variables $\rho $, $\mathbf{v}$ was
established, but it was a one-way connection. In the case of the
irrotational flow the hydrodynamical variables can be expressed via the wave
function $\psi $, but one cannot do this in the case of the irrotational
flow. Hence, one can transit from the description in terms of the wave
function to the description in terms of $\rho $, $\mathbf{v}$, but one
cannot transit from the hydrodynamic description in terms of $\rho $, $%
\mathbf{v}$ to a description in terms of $\psi $, because, in general, the
fluid flow is rotational, and the dynamic system (\ref{a1.22}) cannot be
described in terms of the one-component wave function.

Let us present the wave function in the form 
\begin{equation}
\psi =\sqrt{\rho }e^{i\varphi },  \label{b1.14}
\end{equation}
substitute it in the Schr\"{o}dinger equation (\ref{a1.2a}) and separate the
real and imaginary parts of the equation. We obtain two real equations 
\begin{equation}
\partial _{0}\ln \rho =-\frac{\hbar }{m}\left( \nabla ^{2}\varphi +\nabla
\ln \rho \nabla \varphi \right)  \label{b1.15}
\end{equation}
\begin{equation}
\partial _{0}\varphi +\frac{\hbar }{2m}\left( \nabla \varphi \right) ^{2}=%
\frac{\hbar }{2m}\left( \frac{1}{2}\nabla ^{2}\ln \rho +\left( \frac{1}{2}%
\nabla \ln \rho \right) ^{2}\right)  \label{b1.16}
\end{equation}
To obtain hydrodynamic equation, one needs to take gradient of the equation (%
\ref{b1.16}) and introduce the velocity $\mathbf{v=}\left\{
v^{1},v^{2},v^{3}\right\} $ by means of the relation 
\begin{equation}
\mathbf{v}=\frac{\hbar }{m}\mathbf{\nabla }\varphi  \label{b1.17}
\end{equation}
We obtain 
\begin{equation}
\partial _{0}\rho +\partial _{\alpha }\left( \rho v^{\alpha }\right)
=0,\qquad \partial _{0}v^{\alpha }+v^{\beta }\partial _{\beta }v^{\alpha }%
\mathbf{=-}\frac{1}{\rho }\partial _{\beta }P^{\alpha \beta },\qquad \alpha
=1,2,3  \label{b1.18}
\end{equation}
where $P^{\alpha \beta }$ is the tension tensor 
\begin{equation}
P^{\alpha \beta }=\frac{\hbar ^{2}}{4m^{2}}\left( \frac{\left( \partial
_{\alpha }\rho \right) \partial _{\beta }\rho }{\rho }-\partial _{\alpha
}\partial _{\beta }\rho \right)  \label{b1.18a}
\end{equation}

The hydrodynamic equations (\ref{b1.18}) are obtained as a result of
differentiation of the equation (\ref{a1.2a}), written in terms of the wave
function. It means that to transit from the hydrodynamic equations (\ref%
{b1.18}) to the equation, written in terms of the wave function, one needs
to integrate the hydrodynamic equations (\ref{b1.18}). Besides, in the case
of the irrotational flow the wave function is presented in terms of $\rho $
and hydrodynamical potential $\varphi $. The same is valid in the general
case, but the number of the hydrodynamical potentials is to be more than
one, and it is necessary to introduce additional hydrodynamic (Clebsch)
potentials.

The problem of integration of the hydrodynamical equations is rather
complicated problem, which has been solved only in the end of eighties \cite%
{R89}. To solve this problem, it was necessary to develop a special Jacobian
technique \cite{R99}, which was used already by Clebsch \cite{C57,C59}.

As soon as the hydrodynamic equations for the ideal fluid have been
integrated, it becomes clear, that the wave function is simply a method of
the ideal fluid description. The wave function $\psi $ ceases to be an
enigmatic vector of the Hilbert space, whose meaning was obtained only via
quantum principles. Now one can determine the chain of the dynamic variable
transformations which turn the action (\ref{a1.22}) into the action (\ref%
{b1.5a}) (for details see Appendix A). As a result the action (\ref{a1.22})
may be used as a starting point for the description of the quantum Schr\"{o}%
dinger particle $\mathcal{S}_{\mathrm{S}}$. At such a description the
quantum principles (\ref{b1.5}) are not needed, because they are only a tool
for interpretation of the wave function.

The statistical ensemble (\ref{a1.22}) as the starting point of the quantum
description has a series of advantages over the action (\ref{a1.2}):

\begin{enumerate}
\item The statistical ensemble (\ref{a1.22}) is a very transparent
construction founded on the simple physical idea, that the quantum particle
is a stochastically moving particle.

\item It does not use quantum principles, which are nonrelativistic and
cannot be extended properly to the relativistical case.

\item Statistical ensemble (\ref{a1.22}) is a more general construction,
because the action (\ref{a1.2}) is a partial case of the action (\ref{a1.22}%
).

\item In the statistical description, founded on the action (\ref{a1.22}),
there are three different aspects: dynamical factor, statistical factor and
random factor. Each of these factors can be separated as a corresponding
term in the action and investigated apart.

\item Description in terms of the dynamic system (\ref{a1.22}) is a
statistical description. As any statistical description it contains two
objects: the individual stochastic particle $\mathcal{S}_{\mathrm{st}}$ and
the statistical average particle $\left\langle \mathcal{S}_{\mathrm{st}%
}\right\rangle $. Respectively there are two kinds of measurements:
individual measurement (S-measurement) produced over the individual particle 
$\mathcal{S}_{\mathrm{st}}$ and the massive measurement (M-measurement)
produced over the statistical average particle $\left\langle \mathcal{S}_{%
\mathrm{st}}\right\rangle $. These measurements have different properties,
and their identification is inadmissible.
\end{enumerate}

The complexity of dynamic equations (\ref{b1.11}) is the only defect of the
statistical description (\ref{a1.22}).

We underline that the transition from the action (\ref{a1.22}) as a starting
point to the action (\ref{a1.2}) is motivated mathematically. No additional
physical arguments have been used for the substantiation of the statistical
ensemble (\ref{a1.22}) as a starting point of the quantum description.

If we consider stationary states of the statistical ensemble, we are
interested only in the value of the magnetic moment (which is supposed to be
connected with the value of the total spin). In this case the spin origin is
of no importance. But if we investigate the individual particle structure,
it is important, whether the spin is generated by the individual particle,
or it is generated by vorticity of the fluid flow. It is meaningless to
classify the particles over their spin, if the spin has a collective origin,
and the individual particle has not its own angular moment. If we use the
conventional approach to the quantum mechanics, i.e. if we start from the
action (\ref{a1.2}) we cannot separate dynamical and collective properties
directly. Only starting from the action (\ref{a1.22}), we can try to solve
this problem for concrete dynamic systems (for instance, the Dirac particle $%
\mathcal{S}_{\mathrm{D}}$ and the Pauli particle $\mathcal{S}_{\mathrm{P}})$%
. If a researcher stands on the viewpoint of the Copenhagen interpretation,
where the wave function describes the state of individual particle, the
statement of the problem seems to be incorrect for him.

The collective origin of the spin can be perceived, only using statistical
approach presented by the action (\ref{a1.22}). The statistical description,
founded on the action (\ref{a1.22}) leads to the statement that wave
function describes a state of the statistical ensemble $\mathcal{E}\left[ 
\mathcal{S}_{\mathrm{st}}\right] $, but not a state of a single quantum
particle. Discussion of the question, what object is described by the wave
function, has a long history. Some researchers \cite{H55} believe, that the
wave function describes the state of a single quantum particle, whereas
other ones \cite{B76,B70} believe that the wave function describes the state
of the statistical ensemble. There is a long list of different opinions
about this question, but we do not present them, because this problem is not
a question of a belief. It can and must be solved on the basis of the
mathematical formalism.

The problem is set as follows. What dynamic system is described by the
action (\ref{a1.2})? A single quantum particle, or a statistical ensemble of
single particles? Let us go to the limit $\hbar \rightarrow 0$. Then the
action (\ref{a1.2}) will describe the classical dynamic system $\mathcal{S}_{%
\mathrm{Scl}}$. If the dynamic system $\mathcal{S}_{\mathrm{Scl}}$ is a
single classical particle, then the wave function describes the state of a
single particle. If the dynamic system $\mathcal{S}_{\mathrm{cl}}$ is a
statistical ensemble of classical particles, then the wave function
describes the state of a statistical ensemble of single particles. One
cannot go to the limit $\hbar \rightarrow 0$ in the action (\ref{a1.2})
directly, because the description of the dynamic system $\mathcal{S}_{%
\mathrm{S}}$ degenerates.

We make the change of variables

\begin{equation}
\psi \rightarrow \Psi _{b}=\left\vert \psi \right\vert \exp \left( \frac{%
\hbar }{b}\log \frac{\psi }{\left\vert \psi \right\vert }\right) ,\qquad
\psi =\left\vert \Psi _{b}\right\vert \exp \left( \frac{b}{\hbar }\log \frac{%
\Psi _{b}}{\left\vert \Psi _{b}\right\vert }\right)  \label{a1.9}
\end{equation}
where $b\neq 0$ is some real constant. After this change of variables the
action (\ref{a1.2}) turns into 
\begin{equation*}
\mathcal{S}_{\mathrm{S}}:\qquad \mathcal{A}_{\mathrm{S}}\left[ \Psi
_{b},\Psi _{b}^{\ast }\right] =\int \left\{ \frac{ib}{2}\left( \Psi
_{b}^{\ast }\partial _{0}\Psi _{b}-\partial _{0}\Psi _{b}^{\ast }\cdot \Psi
_{b}\right) -\frac{b^{2}}{2m}\mathbf{\nabla }\Psi _{b}^{\ast }\mathbf{\nabla 
}\Psi _{b}\right.
\end{equation*}
\begin{equation}
-\left. \frac{\hbar ^{2}-b^{2}}{2m}\left( \mathbf{\nabla }\left\vert \Psi
_{b}\right\vert \right) ^{2}\right\} dtd\mathbf{x}  \label{a1.10}
\end{equation}

The dynamic equation takes the form 
\begin{equation}
ib\partial _{0}\Psi _{b}=-\frac{b^{2}}{2m}\mathbf{\nabla }^{2}\Psi _{b}-%
\frac{\hbar ^{2}-b^{2}}{8m}\left( \frac{\left( \mathbf{\nabla }\rho \right)
^{2}}{\rho ^{2}}+2\mathbf{\nabla }\frac{\mathbf{\nabla }\rho }{\rho }\right)
\Psi _{b},\qquad \rho \equiv \Psi _{b}^{\ast }\Psi _{b}  \label{a1.11}
\end{equation}

Instead of (\ref{a1.3}), we obtain 
\begin{equation}
\rho =\Psi _{b}^{\ast }\Psi _{b},\qquad \mathbf{j}=-\frac{ib}{2m}\left( \Psi
_{b}^{\ast }\mathbf{\nabla }\Psi _{b}-\mathbf{\nabla }\Psi _{b}^{\ast }\cdot
\Psi _{b}\right)  \label{a1.12}
\end{equation}

Setting $\hbar =0$ in (\ref{a1.10}), (\ref{a1.11}), we obtain 
\begin{equation*}
\mathcal{S}_{\mathrm{Scl}}:\qquad \mathcal{A}_{\mathrm{Scl}}\left[ \Psi
_{b},\Psi _{b}^{\ast }\right] =\int \left\{ \frac{ib}{2}\left( \Psi
_{b}^{\ast }\partial _{0}\Psi _{b}-\partial _{0}\Psi _{b}^{\ast }\cdot \Psi
_{b}\right) -\frac{b^{2}}{2m}\mathbf{\nabla }\Psi _{b}^{\ast }\mathbf{\nabla 
}\Psi _{b}\right.
\end{equation*}
\begin{equation}
+\left. \frac{b^{2}}{2m}\left( \mathbf{\nabla }\left\vert \Psi
_{b}\right\vert \right) ^{2}\right\} dtd\mathbf{x}  \label{b1.19}
\end{equation}
\begin{equation}
ib\partial _{0}\Psi _{b}=-\frac{b^{2}}{2m}\mathbf{\nabla }^{2}\Psi _{b}+%
\frac{b^{2}}{8m}\left( \frac{\left( \mathbf{\nabla }\rho \right) ^{2}}{\rho
^{2}}+2\mathbf{\nabla }\frac{\mathbf{\nabla }\rho }{\rho }\right) \Psi
_{b},\qquad \rho \equiv \Psi _{b}^{\ast }\Psi _{b}  \label{b1.20}
\end{equation}

The action (\ref{b1.19}) describes the statistical ensemble of free
classical particles and, hence, the wave function describes the statistical
ensemble, but not a single particle. The action (\ref{b1.19}) may not
describe a single classical particle, because the dynamic system (\ref{b1.19}%
) has infinite number of the freedom degrees. As far as the description (\ref%
{b1.19}) in terms of the wave function $\Psi _{b}$ is a limit $\hbar
\rightarrow 0$ of the description in terms of the wave function $\psi $, the
wave function $\psi $ in (\ref{a1.2}) may not describe a single quantum
particle. Thus, the \textit{supposition that the wave function describes a
state of a single particle is incompatible with the quantum mechanics
formalism}.

According to the Copenhagen interpretation of quantum mechanics the wave
function $\psi $ describes the state of a single quantum particle, whereas
the state of a classical particle is described by its position $\mathbf{x}$
and its momentum $\mathbf{p}$. It is supposed that the wave function is a
specific quantum quantity, which has no classical analog. In accordance with
this approach one may not go to the limit $\hbar \rightarrow 0$ in the
action (\ref{a1.2}), because the action vanishes, and the description
degenerates.

The transformation (\ref{a1.9}) changes only the scale of the wave function
phase ln$\left( \psi /\left\vert \psi \right\vert \right) $, and this change
may be very slight. The wave function $\Psi _{b}$ is the valid wave
function, which can be used, in particular, for calculation of average
values by means of the relation (\ref{b1.5}). This calculation may be
produced for any value of the constant $b$. The wave function $\Psi _{b}$
describes the same state of $\mathcal{S}_{\mathrm{S}}$ at different values
of the parameter $b$, because the state of the dynamic system does not
determine the wave function uniquely, and the same state of $\mathcal{S}_{%
\mathrm{S}}$ may be described by different wave functions. From viewpoint of
the statistical description (\ref{a1.22}) the wave function is not uniquely
defined, because it is constructed of hydrodynamic potentials, i.e. it is a
result of integration of uniquely defined velocity $\mathbf{v}$. The
parameter $b$ in the transformation (\ref{b1.5}) is a constant of
integration.

We may set $b=\hbar $ in the relations (\ref{b1.19}), (\ref{b1.20}) and
obtain a description of "classical particle " in the form containing the
quantum constant $\hbar $. 
\begin{equation*}
\mathcal{S}_{\mathrm{Scl}}:\qquad \mathcal{A}_{\mathrm{Scl}}\left[ \psi
,\psi ^{\ast }\right] =\int \left\{ \frac{i\hbar }{2}\left( \psi ^{\ast
}\partial _{0}\psi -\partial _{0}\psi ^{\ast }\cdot \psi \right) -\frac{%
\hbar ^{2}}{2m}\mathbf{\nabla }\psi ^{\ast }\mathbf{\nabla }\psi \right.
\end{equation*}
\begin{equation}
+\left. \frac{\hbar ^{2}}{2m}\left( \mathbf{\nabla }\left\vert \psi
\right\vert \right) ^{2}\right\} dtd\mathbf{x}  \label{b1.21}
\end{equation}
\begin{equation}
i\hbar \partial _{0}\psi =-\frac{\hbar ^{2}}{2m}\mathbf{\nabla }^{2}\psi +%
\frac{\hbar ^{2}}{8m}\left( \frac{\left( \mathbf{\nabla }\rho \right) ^{2}}{%
\rho ^{2}}+2\mathbf{\nabla }\frac{\mathbf{\nabla }\rho }{\rho }\right) \psi
,\qquad \rho \equiv \psi ^{\ast }\psi  \label{b1.22}
\end{equation}
The same result may be obtained from (\ref{b1.19}), (\ref{b1.20}) by means
of the transformation inverse to the transformation (\ref{a1.9}). Formally
the action (\ref{b1.21}) distinguishes from the action (\ref{a1.2}) in the
last term, which describes a lack of quantum effects. The quantum constant
in two first terms has no relation to quantum effects. The dependence on $%
\hbar $ is conditioned by a special choice of the arbitrary constant $b$.

The action (\ref{b1.21}) describes the dynamic system $\mathcal{S}_{\mathrm{%
Scl}}=\mathcal{E}\left[ \mathcal{S}_{\mathrm{d}}\right] $ in the "quantum
language", i.e. in terms of the wave function. The action 
\begin{equation}
\mathcal{S}_{\mathrm{Scl}}=\mathcal{E}\left[ \mathcal{S}_{\mathrm{d}}\right]
:\qquad \mathcal{A}_{\mathcal{E}\left[ \mathcal{S}_{\mathrm{d}}\right] }%
\left[ \mathbf{x}\right] =\int \frac{m}{2}\left( \frac{d\mathbf{x}}{dt}%
\right) ^{2}dtd\mathbf{\xi }  \label{b1.23}
\end{equation}
where $\mathbf{x}=\mathbf{x}\left( t,\mathbf{\xi }\right) $, describes the
same dynamic system in the "classical language", i.e. in terms of classical
variables $\mathbf{x},\mathbf{p}$. In the same way the action (\ref{a1.2})
describes the dynamic system $\mathcal{S}_{\mathrm{S}}=\mathcal{E}\left[ 
\mathcal{S}_{\mathrm{st}}\right] $ in quantum language, whereas the action (%
\ref{a1.22}) describes the same dynamic system in the classical language. It
is reasonable that the quantum system $\mathcal{S}_{\mathrm{S}}$ is
described simpler in the quantum language, whereas the classical system $%
\mathcal{S}_{\mathrm{Scl}}=\mathcal{E}\left[ \mathcal{S}_{\mathrm{d}}\right] 
$ is described simpler in the classical language. However, it is not a
reason for the statement that the quantum system is to be described in the
quantum language (in terms of the wave function).

Two different description of the classical system $\mathcal{S}_{\mathrm{cl}}$
can be used for interpretation of the rule (\ref{b1.5}) and for
interpretation of the correspondence principle. The obtained results may be
applied to the quantum system $\mathcal{S}_{\mathrm{S}}$, because the
difference between the dynamic systems $\mathcal{S}_{\mathrm{S}}=\mathcal{E}%
\left[ \mathcal{S}_{\mathrm{st}}\right] $ and $\mathcal{S}_{\mathrm{Scl}}=%
\mathcal{E}\left[ \mathcal{S}_{\mathrm{d}}\right] $, described respectively
by actions (\ref{a1.2}) and (\ref{b1.21}), manifests itself only in the
additional nonlinear term in the dynamic equation. The possibility of
description $\mathcal{S}_{\mathrm{S}}=\mathcal{E}\left[ \mathcal{S}_{\mathrm{%
st}}\right] $ and $\mathcal{S}_{\mathrm{Scl}}=\mathcal{E}\left[ \mathcal{S}_{%
\mathrm{d}}\right] $ in both languages (classical and quantum) shuts the
door before the Copenhagen interpretation, where the wave function is
supposed to describe a single particle. Thus, \textit{there is neither
reason nor excuse for application of the Copenhagen interpretation}.

The rules (\ref{b1.5}) are statistical relations, which can be applied to
both classical and quantum statistical ensembles. Some results of their
application appear to be rather curious. For instance, the momentum
distribution 
\begin{equation}
w\left( \mathbf{p}\right) =\psi _{\mathbf{p}}^{\ast }\psi _{\mathbf{p}%
},\qquad \psi _{\mathbf{p}}=\frac{1}{\left( 2\pi \right) ^{3}}\int e^{\frac{i%
}{\hbar }\mathbf{px}}\psi \left( \mathbf{x}\right) d\mathbf{x}  \label{b1.24}
\end{equation}
at the state described by the wave function $\psi $ appears to be rather the
mean momentum distribution, than the momentum distribution \cite{R2004b}.
Let us manifest the difference between the momentum distribution and the
mean momentum distribution (distribution over mean momenta) in the example
of the ideal gas.

Let us consider a gas, moving with the constant velocity $\mathbf{u}\left( 
\mathbf{x}\right) =\mathbf{u}=$const. As any fluid such a gas motion may be
described by the wave function.\ It has the form 
\begin{equation}
\psi \left( t,\mathbf{x}\right) =A\exp \left( -\frac{im\mathbf{ux}}{\hbar }-%
\frac{im\mathbf{u}^{2}}{2\hbar }t\right) ,\qquad A=\text{const}
\label{b1.25}
\end{equation}%
where $m$ is the mass of the gas molecule. The density $\rho $ and the
velocity $\mathbf{u}$, described by the formulas (\ref{a1.3}), (\ref{b1.25}) 
\begin{equation}
\rho =\psi ^{\ast }\psi ,\qquad \mathbf{u}=\frac{\mathbf{j}}{\rho }=-\frac{%
i\hbar }{2m\psi ^{\ast }\psi }\left( \psi ^{\ast }\mathbf{\nabla }\psi -%
\mathbf{\nabla }\psi ^{\ast }\cdot \psi \right)  \label{b1.25a}
\end{equation}%
are constant and satisfy the hydrodynamic equations with arbitrary form of
the internal energy.

Calculation by means of the formula (\ref{b1.24}) gives 
\begin{equation}
w\left( \mathbf{p}\right) d\mathbf{p}=\psi _{\mathbf{p}}^{\ast }\psi _{%
\mathbf{p}}d\mathbf{p}=B\delta \left( \mathbf{p}-m\mathbf{u}\right) d\mathbf{%
p}  \label{b1.26}
\end{equation}
where $B$ is a constant and $\delta $ is the Dirac $\delta $-function.
Chaotic motion of molecules is described by the Maxwell distribution 
\begin{equation}
F\left( \mathbf{x},\mathbf{p}\right) d\mathbf{p}=\frac{1}{\left( 2\pi
mkT\right) ^{3/2}}\exp \left\{ -\frac{\left( \mathbf{p}-m\mathbf{u}\left( 
\mathbf{x}\right) \right) ^{2}}{2mkT}\right\} d\mathbf{p}  \label{b1.27}
\end{equation}
It depends on the gas temperature $T$ and has nothing to do with the
distribution (\ref{b1.26}). The gas motion is described by the gas dynamic
equations, which do not take into account chaotic molecular motion and do
not contain a reference to the Maxwell distribution. What is the
distribution (\ref{b1.26})?

Let us divide the volume $V$ of the gas flow into similar cubic cells $%
V_{1},V_{2},...V_{N}$, $N\gg 1$. Let the following conditions be satisfied 
\begin{equation}
l_{\mathrm{c}}\ll L,\qquad \left\vert v_{\mathrm{t}}\tau _{c}\right\vert \ll
L,\qquad \left\vert \mathbf{u}\left( \mathbf{x}\right) \right\vert \ll v_{%
\mathrm{t}}=\sqrt{\frac{3kT}{m}}  \label{b1.28}
\end{equation}
where $L$ is the linear size of the cell, $l_{\mathrm{c}}$ is the mean path
between the molecule collisions, $\tau _{\mathrm{c}}$ is the mean time
between the collisions and $v_{\mathrm{t}}$ is the mean thermal velocity of
molecules.

Let us calculate the mean momentum $\left\langle \mathbf{p}_{i}\right\rangle 
$ of the gas molecule in the cell $V_{i}$. We obtain $\left\langle \mathbf{p}%
_{i}\right\rangle =m\mathbf{u}\left( \mathbf{x}\right) $, \ $\mathbf{x}\in
V_{i}$, $i=1,2,...N$. The set of all $\left\langle \mathbf{p}%
_{i}\right\rangle $, $i=1,2,...N$ forms the mean momentum distribution. This
distribution is determined completely by the gas flow, and it has nothing to
do with the Maxwell momentum distribution (\ref{b1.27}), which describes
both the regular and random components of the molecule momenta. Under
conditions (\ref{b1.28}) the mean momentum distribution is much narrower,
than the Maxwell distribution, because the Maxwell distribution takes into
account the random component of the molecule velocity, and in the given case
the random component is much larger, than the regular one. In the given case
the relation (\ref{b1.26}) may be rewritten in the form 
\begin{equation}
w\left( \left\langle \mathbf{p}\right\rangle \right) d\left\langle \mathbf{p}%
\right\rangle =\psi _{\mathbf{p}}^{\ast }\psi _{\mathbf{p}}d\left\langle 
\mathbf{p}\right\rangle =B\delta \left( \left\langle \mathbf{p}\right\rangle
-m\mathbf{u}\right) d\left\langle \mathbf{p}\right\rangle  \label{b1.29}
\end{equation}
where $\left\langle \mathbf{p}\right\rangle $ is the mean particle momentum.

Besides, any $\left\langle \mathbf{p}_{i}\right\rangle $ is labelled by the
index $i$, or by the coordinate $\mathbf{x}_{i}$ of the volume $V_{i}$. It
means, that variables $\mathbf{x}$ and $\left\langle \mathbf{p}%
_{i}\right\rangle $ are not independent, and mutual coordinate-momentum
distribution does not exist. In the Copenhagen interpretation the lack of
the mutual coordinate-momentum distribution is explained by the
noncommutativity of operators $\mathbf{x}$ and $\mathbf{p}=-i\hbar \nabla $,
and the distribution (\ref{b1.24}) is considered to be a distribution over
the stochastic component of the momentum (some quantum analog of the Maxwell
distribution). There are other unexpected characteristics of the rule (\ref%
{b1.5}).

The rule (\ref{b1.5}) is only a method to obtain the information contained
in the investigated dynamic system. This information can be obtained from
the dynamic system by other methods. An application of the rule (\ref{b1.5})
does not add any real information beyond that one, which is contained in the
investigated dynamic system.

\section{Dynamic disquantization}

The quantum langauge, i.e. the description, containing the quantum constant $%
\hbar $, may be used for a description of a classical dynamic system,
because the quantum constant $\hbar $ may be used instead of the arbitrary
dynamical constant $b$. Replacement of dynamical constant $b$ by the quantum
constant is produced to make the dynamic equations to be linear. For
instance, in the action (\ref{a1.10}) the quantum constant $\hbar $ is used
naturally, i.e. in the sense that setting $\hbar =0$, we suppress the
quantum effects. In the action (\ref{a1.2}) for the same dynamic system the
quantum constant $\hbar $ is used artificially in the sense that setting $%
\hbar =0$, we do not suppress the quantum effects. Furthermore, setting $%
\hbar =0$, we destroy any description. But the action (\ref{a1.2}) generates
linear dynamic equation, and this circumstance is a reason of the artificial
identification $b=\hbar $, when the dynamical constant $b$ is identified
with the quantum constant $\hbar $.

Such an artificial identification may be produced in other quantum systems
(for instance, in $\mathcal{S}_{\mathrm{D}}$ and $\mathcal{S}_{\mathrm{P}}$%
), and we cannot be sure, that setting $\hbar =0$, we suppress the quantum
effects. Besides, we cannot be sure that, using the transformation of the
type (\ref{a1.9}), we can separate the quantum terms from dynamical and
statistical ones.

We need a more effective formal dynamical procedure, which could suppress
the stochastic terms. Let us compare dynamic equations (\ref{b1.11}) for $%
\mathcal{S}_{\mathrm{S}}=\mathcal{E}\left[ \mathcal{S}_{\mathrm{st}}\right] $
written in the form 
\begin{equation}
\frac{d\mathbf{p}}{dt}=-\mathbf{\nabla }U\left( \rho ,\mathbf{\nabla }\rho
\right) ,\qquad \frac{d\mathbf{x}}{dt}=\frac{\mathbf{p}}{m},\qquad U\left(
\rho ,\mathbf{\nabla }\rho \right) =\frac{\hbar ^{2}}{8m}\left( \frac{\left( 
\mathbf{\nabla }\rho \right) ^{2}}{\rho ^{2}}-2\frac{\mathbf{\nabla }%
^{2}\rho }{\rho }\right)  \label{b2.2}
\end{equation}%
with the dynamic equations for $\mathcal{S}_{\mathrm{Scl}}=\mathcal{E}\left[ 
\mathcal{S}_{\mathrm{d}}\right] $, which have the form 
\begin{equation}
\frac{d\mathbf{p}}{dt}=0,\qquad \frac{d\mathbf{x}}{dt}=\frac{\mathbf{p}}{m}
\label{b2.1}
\end{equation}%
where $\mathbf{x}=\mathbf{x}\left( t,\mathbf{\xi }\right) $, $\mathbf{p}=%
\mathbf{p}\left( t,\mathbf{\xi }\right) $. Dynamic equations (\ref{b2.2}),
are the partial differential equations, because $\rho $ is defined by the
relation (\ref{b1.9}), containing derivatives with respect to $\xi _{\alpha
} $, $\alpha =1,2,3$, whereas dynamic equations (\ref{b2.1}) are ordinary
differential equations. Equations (\ref{b2.1}) contain derivatives only in
one direction in the space of independent variables $\left\{ t,\mathbf{\xi }%
\right\} $, whereas equation (\ref{b2.2}) contain derivatives in different
directions of the space of independent variables $\left\{ t,\mathbf{\xi }%
\right\} $. This property is conserved at any change of independent
dynamical variables, and, in particular, at the change $\left\{ t,\mathbf{%
\xi }\right\} \rightarrow \left\{ t,\mathbf{x}\right\} $. If a system of
partial differential equations contains derivative only in one direction of
the space of independent variables, this system can be reduced to the system
of ordinary differential equations by means of a proper change of variables.

If we want to suppress the quantum effects, we must to reduce the system of
partial differential equations to the system of ordinary differential
equations. To make this, we should project derivatives in the space of
independent variables onto some direction. Then the system will contain
derivatives only in one direction, and hence it may be reduced to the system
of ordinary differential equations. Onto what direction should derivatives
in the system (\ref{b2.2}) be projected, to obtain the system (\ref{b2.1})?

Such a direction is described by the 4-current $j^{k}=\left\{ \rho ,\mathbf{j%
}\right\} =\left\{ j^{k}\right\} ,$ $k=0,1,2,3$ in the space-time. The
projection should be made in the space of independent variables $\left\{ t,%
\mathbf{x}\right\} $, i.e. in the space-time. It is convenient to choose
dependent variables in such a way, that the 4-current $j^{k}$ were one of
dependent variables. We take the action (\ref{A.12}) for the dynamic system $%
\mathcal{S}_{\mathrm{S}}=\mathcal{E}\left[ \mathcal{S}_{\mathrm{st}}\right] $%
\begin{equation}
\mathcal{A}_{\mathcal{E}\left[ \mathcal{S}_{\mathrm{st}}\right] }\left[
\varphi ,\mathbf{\xi },j\right] =\int \left\{ \frac{m}{2}\frac{j^{\alpha
}j^{\alpha }}{\rho }-bj^{k}\left( \partial _{k}\varphi +g^{\alpha }\left( 
\mathbf{\xi }\right) \partial _{k}\xi _{\alpha }\right) -\frac{\hbar ^{2}}{8m%
}\frac{\left( \mathbf{\nabla }\rho \right) ^{2}}{\rho }\right\} d^{4}x%
\mathbf{,}  \label{b2.3}
\end{equation}%
where according to (\ref{A.17}) and (\ref{g1.30}) 
\begin{equation}
j^{k}=\left\{ \rho ,\mathbf{j}\right\} =\left\{ \rho ,\frac{b\rho }{m}\left( 
\mathbf{\nabla }\varphi +g^{\alpha }\left( \mathbf{\xi }\right) \mathbf{%
\nabla }\xi _{\alpha }\right) \right\}  \label{b2.4}
\end{equation}%
and $g^{\alpha }\left( \mathbf{\xi }\right) $, $\alpha =1,2,3$ are arbitrary
functions of argument $\mathbf{\xi }$.

The second term in the action (\ref{b2.3}) contains derivatives only in the
direction of the 4-vector $j^{k}$. In the last term of (\ref{b2.3}) the
derivatives are to be projected onto the vector $j^{k}$. We are to make the
change 
\begin{equation}
\partial _{l}\rightarrow \partial _{||l}=\frac{j_{l}j^{k}}{j_{s}j^{s}}%
\partial _{k},\qquad l=0,1,2,3  \label{b2.5}
\end{equation}
in the action (\ref{b2.3}). We obtain 
\begin{equation}
\frac{\left( \mathbf{\nabla }\rho \right) ^{2}}{\rho }=\frac{\left( \mathbf{%
\partial }_{\alpha }\rho \right) \left( \mathbf{\partial }_{\alpha }\rho
\right) }{\rho }\rightarrow \frac{j_{\alpha }j_{\alpha }\left( j^{i}\partial
_{i}\rho \right) ^{2}}{\rho \left( j^{s}j_{s}\right) ^{2}}  \label{b2.6}
\end{equation}
\begin{equation*}
j_{\alpha }j_{\alpha }=\mathbf{j}^{2}=\rho ^{2}\mathbf{v}^{2},\qquad
j^{s}j_{s}=c^{2}\rho ^{2}-\rho ^{2}\mathbf{v}^{2}
\end{equation*}
In the nonrelativistic approximation, when the velocity $\left\vert \mathbf{v%
}\right\vert \ll c$, we obtain the following estimation 
\begin{equation}
\frac{\left( \mathbf{\nabla }\rho \right) ^{2}}{\rho }\approx \frac{\mathbf{v%
}^{2}\left( j^{i}\partial _{i}\rho \right) ^{2}}{c^{4}\rho ^{3}}
\label{b2.7}
\end{equation}
In the nonrelativistic approximation $c\rightarrow \infty $ the last term in
the action (\ref{b2.3}) is to be neglected after the change (\ref{b2.5}).
Thus, in the case of the Schr\"{o}dinger particle $\mathcal{S}_{\mathrm{S}}$
the change (\ref{b2.5}) leads to a suppression of quantum effects.

We shall refer to the procedure (\ref{b2.5}) as the dynamical
disquantization, because it transforms the Schr\"{o}dinger particle $%
\mathcal{S}_{\mathrm{S}}=\mathcal{E}\left[ \mathcal{S}_{\mathrm{st}}\right] $
into the classical system $\mathcal{S}_{\mathrm{Scl}}=\mathcal{E}\left[ 
\mathcal{S}_{\mathrm{d}}\right] $. The dynamical disquantization is the
relativistic dynamical procedure. It does not refer to the quantum constant
and suppresses any stochasticity regardless of its origin. From here on we
shall use the dynamical disquantization for the suppression of stochasticity
in quantum systems.

Strictly, the dynamical disquantization is to be applied to the dynamic
equations. But in many cases the application of the dynamical
disquantization to the action leads to the same result, as its application
to the dynamic equations.

\section{Nonrelativistic Dirac particle}

The Dirac particle is the dynamic system $\mathcal{S}_{\mathrm{D}}$,
described by the Dirac equation. The action $\mathcal{A}_{\mathrm{D}}$ for
the dynamic system $\mathcal{S}_{\mathrm{D}}$ has the form 
\begin{equation}
\mathcal{S}_{\mathrm{D}}:\qquad \mathcal{A}_{\mathrm{D}}[\bar{\psi},\psi
]=c^{2}\int (-mc\bar{\psi}\psi +\frac{i}{2}\hbar \bar{\psi}\gamma
^{l}\partial _{l}\psi -\frac{i}{2}\hbar \partial _{l}\bar{\psi}\gamma
^{l}\psi -\frac{e}{c}A_{l}\bar{\psi}\gamma ^{l}\psi )d^{4}x  \label{b1.1}
\end{equation}%
where $m$ and $e$ are respectively the mass and the charge of the Dirac
particle, and $c$ is the speed of the light. Here $\psi $ is four-component
complex wave function, $\psi ^{\ast }$ is the Hermitian conjugate wave
function, and $\bar{\psi}=\psi ^{\ast }\gamma ^{0}$ is the conjugate one. $%
\gamma ^{i}$, $i=0,1,2,3$ are $4\times 4$ complex constant matrices,
satisfying the relation 
\begin{equation}
\gamma ^{l}\gamma ^{k}+\gamma ^{k}\gamma ^{l}=2g^{kl}I,\qquad k,l=0,1,2,3.
\label{b1.2}
\end{equation}%
where $I$ is the $4\times 4$ identity matrix, and $g^{kl}=$diag$\left(
c^{-2},-1,-1,-1\right) $ is the metric tensor. The quantity $A_{k}$, $%
k=0,1,2,3$ is the electromagnetic potential. The action (\ref{b1.1})
generates dynamic equation for the dynamic system $\mathcal{S}_{\mathrm{D}}$%
, known as the Dirac equation 
\begin{equation}
\gamma ^{l}\left( -i\hbar \partial _{l}+\frac{e}{c}A_{l}\right) \psi +mc\psi
=0  \label{f1.2}
\end{equation}%
and expressions for physical quantities: the 4-flux $j^{k}$ of particles and
the energy-momentum tensor $T_{l}^{k}$%
\begin{equation}
j^{k}=c^{2}\bar{\psi}\gamma ^{k}\psi ,\qquad T_{l}^{k}=\frac{ic^{2}}{2}%
\left( \bar{\psi}\gamma ^{k}\partial _{l}\psi -\partial _{l}\bar{\psi}\cdot
\gamma ^{k}\psi \right)  \label{f1.3}
\end{equation}

Here we obtain nonrelativistic approximation of the Dirac particle. Our
investigation differs from the conventional derivation of this approximation
(see for instance \cite{D58}) by consideration of the high frequency
solutions. To obtain the nonrelativistic approximation $\mathcal{S}_{\mathrm{%
nD}}$ of the Dirac particle $\mathcal{S}_{\mathrm{D}}$, we use the following
representation of the $4\times 4$ Dirac $\gamma $-matrices 
\begin{equation}
\gamma ^{0}=\frac{1}{c}\left( 
\begin{array}{ll}
0 & I \\ 
I & 0%
\end{array}%
\right) ,\qquad \gamma ^{\mu }=\left( 
\begin{array}{ll}
0 & -\sigma _{\mu } \\ 
\sigma _{\mu } & 0%
\end{array}%
\right) ,\qquad \gamma ^{0}\gamma ^{\mu }=\frac{1}{c}\left( 
\begin{array}{ll}
\sigma _{\mu } & 0 \\ 
0 & -\sigma _{\mu }%
\end{array}%
\right)  \label{B.1}
\end{equation}%
where $\sigma _{\mu }$, $\mu =1,2,3$ are $2\times 2$ Pauli matrices, and $I$
is the $2\times 2$ identity matrix. We use designations 
\begin{equation}
\pi _{l}\equiv i\hbar \partial _{l}-\frac{e}{c}A_{l},\qquad \pi _{l}^{\ast
}\equiv -i\hbar \partial _{l}-\frac{e}{c}A_{l},\qquad l=0,1,2,3  \label{B.2}
\end{equation}%
and representation of 4-component wave functions $\psi $ in the form 
\begin{equation}
\psi =\frac{\exp \left( -\frac{i}{2}\Omega t\right) }{\sqrt{2}}\left( 
\begin{array}{l}
\psi _{1}+\psi _{2} \\ 
\psi _{1}-\psi _{2}%
\end{array}%
\right) ,\qquad \psi ^{\ast }=\frac{\exp \left( \frac{i}{2}\Omega t\right) }{%
\sqrt{2}}\left( \psi _{1}^{\ast }+\psi _{2}^{\ast },\psi _{1}^{\ast }-\psi
_{2}^{\ast }\right) ,  \label{B.3}
\end{equation}%
\begin{equation}
\Omega =\frac{2mc^{2}}{\hbar }  \label{B.18}
\end{equation}%
where $\psi _{1}$ and $\psi _{2}$ are two-component wave functions, and
asterisk $\left( ^{\ast }\right) $ means the Hermitian conjugation. The
action (\ref{b1.1}) for the Dirac particle $\mathcal{S}_{\mathrm{D}}$ can be
written in the form%
\begin{equation*}
\mathcal{A}_{\mathrm{D}}\left[ \psi _{1},\psi _{1}^{\ast },\psi _{2},\psi
_{2}^{\ast }\right] =\frac{1}{2}\int \left\{ \psi _{1}^{\ast }\pi _{0}\psi
_{1}+\psi _{2}^{\ast }\pi _{0}\psi _{2}+c\psi _{1}^{\ast }\sigma _{\mu }\pi
_{\mu }\psi _{2}\right.
\end{equation*}
\begin{equation}
\qquad \qquad \qquad \left. +c\psi _{2}^{\ast }\sigma _{\mu }\pi _{\mu }\psi
_{1}+2mc^{2}\psi _{2}^{\ast }\psi _{2}\right\} d^{4}x+\text{c.c.}
\label{B.5}
\end{equation}%
where (c.c.) means the complex conjugate expression with respect to the
previous term.

Dynamic equations have the form 
\begin{eqnarray}
\delta \psi _{1}^{\ast } &:&\qquad \pi _{0}\psi _{1}=-c\sigma _{\mu }\pi
_{\mu }\psi _{2}  \label{B.6} \\
\delta \psi _{2}^{\ast } &:&\qquad \left( \pi _{0}+2mc^{2}\right) \psi
_{2}=-c\sigma _{\mu }\pi _{\mu }\psi _{1}  \label{B.7}
\end{eqnarray}%
Expressions (\ref{f1.3}) for the 4-current $j^{k}$ and the energy-momentum
tensor $T_{k}^{0}$ turn into 
\begin{equation}
j^{0}=\left( \psi _{1}^{\ast }\psi _{1}+\psi _{2}^{\ast }\psi _{2}\right)
,\qquad j^{\mu }=c\left( \psi _{1}^{\ast }\sigma _{\mu }\psi _{2}+\psi
_{2}^{\ast }\sigma _{\mu }\psi _{1}\right) ,\qquad \mu =1,2,3  \label{B.7c}
\end{equation}

\begin{equation}
T_{0}^{0}=2mc^{2}\left( \psi _{1}^{\ast }\psi _{1}+\psi _{2}^{\ast }\psi
_{2}\right) +\frac{i\hbar }{2}\left( \psi _{1}^{\ast }\partial _{0}\psi
_{1}+\psi _{2}^{\ast }\partial _{0}\psi _{2}\right) +\text{c.c.}
\label{B.7d}
\end{equation}%
\begin{equation}
T_{\mu }^{0}=\frac{i\hbar }{2}\left( \psi _{1}^{\ast }\partial _{\mu }\psi
_{1}+\psi _{2}^{\ast }\partial _{\mu }\psi _{2}\right) +\text{c.c.},\qquad
\mu =1,2,3  \label{B.7e}
\end{equation}%
\begin{equation*}
T_{0}^{\nu }=2mc^{2}\left( \psi _{1}^{\ast }\sigma _{\nu }\psi _{2}+\psi
_{2}^{\ast }\sigma _{\nu }\psi _{1}\right) +\frac{i\hbar }{2}\left( \psi
_{1}^{\ast }\sigma _{\nu }\partial _{0}\psi _{2}+\psi _{2}^{\ast }\sigma
_{\nu }\partial _{0}\psi _{1}\right) +\text{c.c.},\qquad \nu =1,2,3
\end{equation*}%
In the nonrelativistic case, when the speed of the light $c\rightarrow
\infty $, we have $\left\vert \pi _{\mu }\psi _{1}\right\vert \ll
mc\left\vert \psi _{1}\right\vert $, and according to (\ref{B.7}) $\psi _{2}$
is a small quantity, provided the temporal frequency of the quantity $\psi
_{2}$ is not too large. For simplicity we shall consider the case, when $%
A_{0}=0$ and $\pi _{0}=i\hbar \partial _{0}$. In this case, resolving
equations (\ref{B.6}) and (\ref{B.7}) with respect to $\psi _{1}$, we obtain 
\begin{equation}
\left( i\hbar \partial _{0}+2mc^{2}\right) i\hbar \partial _{0}\psi
_{1}=c^{2}\pi _{\nu }\sigma _{\nu }\sigma _{\mu }\pi _{\mu }\psi _{1}
\label{B.7a}
\end{equation}%
Using identity 
\begin{equation}
\sigma _{\mu }\sigma _{\nu }\equiv \delta _{\mu \nu }I+i\varepsilon _{\mu
\nu \alpha }\sigma _{\alpha },\qquad \mu ,\nu =1,2,3  \label{B.7f}
\end{equation}%
where $I$ is the $2\times 2$ identity matrix and $\varepsilon _{\mu \nu
\alpha }$ is the Levi-Chivita pseudotensor, we can transform the dynamic
equation (\ref{B.7a}) to the form 
\begin{equation}
\left( \frac{i}{\Omega }\partial _{0}+1\right) i\hbar \partial _{0}\psi _{1}=%
\hat{H}_{\mathrm{P}}\left( m\right) \psi _{1},\qquad \Omega =\frac{2mc^{2}}{%
\hbar }  \label{B.7b}
\end{equation}%
Here $\Omega $ is the threshold frequency (\ref{B.18}) of the pair
production, and $\hat{H}_{\mathrm{P}}\left( m\right) $ is the Hamiltonian
for the Pauli equation. It has the form 
\begin{equation}
\hat{H}_{\mathrm{P}}\left( m\right) =\frac{\pi _{\mu }\pi _{\mu }}{2m}+\frac{%
ie\hbar }{2mc}\varepsilon _{\nu \mu \alpha }\partial _{\nu }A_{\mu }\sigma
_{\alpha }=\frac{\mathbf{\pi }^{2}}{2m}+\frac{ie\hbar }{2mc}\mathbf{H\sigma }
\label{B.8}
\end{equation}%
where $\mathbf{\pi }=\left\{ \pi _{1},\pi _{2},\pi _{3}\right\} $, $\mathbf{%
\sigma }=\left\{ \sigma _{1},\sigma _{2},\sigma _{3}\right\} $ and $\mathbf{H%
}=\left\{ H_{1},H_{2},H_{3}\right\} $ is the magnetic field 
\begin{equation}
H_{\alpha }=\varepsilon _{\nu \mu \alpha }\partial _{\nu }A_{\mu },\qquad
\alpha =1,2,3  \label{B.17}
\end{equation}

Let $u_{\pm }$ be the low frequency $\left( \left\vert \partial
_{0}u\right\vert \ll \Omega \left\vert u\right\vert \right) $ solutions of
the Pauli equations 
\begin{equation}
i\hbar \partial _{0}u_{+}=\hat{H}_{\mathrm{P}}\left( m\right) u_{+},\qquad
i\hbar \partial _{0}u_{-}=\hat{H}_{\mathrm{P}}\left( -m\right) u_{-}
\label{B.17a}
\end{equation}
In the nonrelativistic case, when $\left\vert \mathbf{\pi }u_{\pm
}\right\vert \ll \left\vert mcu_{\pm }\right\vert $, the solution $u_{\pm }$
of equations (\ref{B.17a}) is a low frequency solution, i.e. $\left\vert
\partial _{0}u_{\pm }\right\vert \ll \Omega \left\vert u_{\pm }\right\vert $%
, as it follows from (\ref{B.17a}). The low frequency solution $u_{+}\left(
t,\mathbf{x}\right) $ is a solution of the equation (\ref{B.7b}), because
the first term in lhs of (\ref{B.7b}) is small as compared with the second
one.

Let us consider the case, when $\psi _{1}=u_{+}\left( t,\mathbf{x}\right) $
is the low frequency solution of the first equation (\ref{B.17a}). In this
case the first equation (\ref{B.17a}) coincides with (\ref{B.7a}). It
follows from (\ref{B.7}), that 
\begin{equation}
\psi _{2}=-\frac{\sigma _{\mu }\pi _{\mu }}{2mc}\psi _{1}  \label{B.18a}
\end{equation}%
and $\left\vert \psi _{2}\right\vert \ll \left\vert \psi _{1}\right\vert $,
because in the nonrelativistic approximation $\left\vert \pi _{\mu }\psi
_{1}\right\vert \ll mc\left\vert \psi _{1}\right\vert $. It follows from (%
\ref{B.7c}) 
\begin{equation}
j^{0}=\psi _{1}^{\ast }\psi _{1}  \label{B.19a}
\end{equation}%
\begin{equation*}
j^{\mu }=-\frac{1}{2m}\left( \psi _{1}^{\ast }\pi _{\mu }\psi _{1}+\left(
\pi _{\mu }^{\ast }\psi _{1}^{\ast }\right) \psi _{1}+i\varepsilon _{\mu \nu
\alpha }\psi _{1}^{\ast }\sigma _{\alpha }\pi _{\nu }\psi _{1}-i\varepsilon
_{\mu \nu \alpha }\left( \pi _{\nu }^{\ast }\psi _{1}^{\ast }\right) \sigma
_{\alpha }\psi _{1}\right)
\end{equation*}%
or 
\begin{equation}
\mathbf{j}=-\frac{i\hbar }{2m}\left( \psi _{1}^{\ast }\mathbf{\nabla }\psi
_{1}-\mathbf{\nabla }\psi _{1}^{\ast }\cdot \psi _{1}\right) +\frac{e}{c}%
\mathbf{A}\psi _{1}^{\ast }\psi _{1}+\frac{\hbar }{2m}\mathbf{\nabla \times }%
\left( \psi _{1}^{\ast }\mathbf{\sigma }\psi _{1}\right)  \label{B.19b}
\end{equation}

The high frequency expression $\exp \left( i\Omega t\right) u_{-}\left( t,%
\mathbf{x}\right) $ is also a solution of (\ref{B.7b}). Indeed, substituting
it in (\ref{B.7b}), we obtain after transformation of lhs 
\begin{equation}
e^{i\Omega t}\left( \frac{i}{\Omega }\partial _{0}-1\right) i\hbar \partial
_{0}u_{-}\left( t,\mathbf{x}\right) =-e^{i\Omega t}\hat{H}_{\mathrm{P}%
}\left( -m\right) u_{-}\left( t,\mathbf{x}\right)  \label{B.19}
\end{equation}%
As far as $u_{-}\left( t,\mathbf{x}\right) $ is a low frequency quantity $%
\left( \left\vert \partial _{0}u_{-}\right\vert \ll \Omega \left\vert
u_{-}\right\vert \right) $, the first term in lhs of (\ref{B.19}) is small
as compared with the second one, and the function $u_{-}\left( t,\mathbf{x}%
\right) $ is a solution of the second equation (\ref{B.17a}). Thus, the
general solution of (\ref{B.7b}) has the form 
\begin{equation}
\psi _{1}=u_{+}\left( t,\mathbf{x}\right) +e^{i\Omega t}u_{-}\left( t,%
\mathbf{x}\right)  \label{B.20}
\end{equation}%
where $u_{+}\left( t,\mathbf{x}\right) $ and $u_{-}\left( t,\mathbf{x}%
\right) $ are two independent low frequency solutions of (\ref{B.17a}).
Equations for $u_{+}\left( t,\mathbf{x}\right) $ and $u_{-}\left( t,\mathbf{x%
}\right) $ are the Pauli equations with different sign of the mass $m$.

The quantity $\psi _{2}$ is determined by the equation (\ref{B.7}), whose
general solution has the form 
\begin{equation}
\psi _{2}=\exp \left( i\Omega t\right) \left( w_{0}-\dint \frac{c\pi _{\mu
}\sigma _{\mu }}{i\hbar }\psi _{1}\left( t,\mathbf{x}\right) \exp \left(
-i\Omega t\right) dt\right)  \label{B.21}
\end{equation}%
where $\Omega $ is determined by the relation (\ref{B.18}), and $w_{0}$ is
an indefinite constant, which can be included in the indefinite integral in (%
\ref{B.21}). (The quantity $w_{0}$ is a constant, but not a function of $%
\mathbf{x}$, because the quantity $\psi _{2}$ is to satisfy equation (\ref%
{B.6}), and it is possible only if $w_{0}=$const). In the general case, when 
$\psi _{1}$ has the form (\ref{B.20}), we obtain for $\psi _{2}$ 
\begin{equation}
\psi _{2}=w_{0}e^{i\Omega t}-e^{i\Omega t}\frac{c\pi _{\mu }\sigma _{\mu }}{%
i\hbar }\int \left( u_{+}\left( t,\mathbf{x}\right) e^{-i\Omega
t}+u_{-}\left( t,\mathbf{x}\right) \right) dt  \label{B.22}
\end{equation}%
In the limit $\Omega \rightarrow \infty $, the first term in the integral of
(\ref{B.22}) is small as compared with the second term (if $u_{-}\left( t,%
\mathbf{x}\right) \neq 0$). Integrating the first term in integral of (\ref%
{B.22}), we consider $u_{+}\left( t,\mathbf{x}\right) $ as independent of $t$%
. We obtain 
\begin{equation}
\psi _{2}\left( t,\mathbf{x}\right) =-\frac{c\sigma _{\mu }\pi _{\mu }}{%
i\hbar }u\left( t,\mathbf{x}\right) e^{i\Omega t}-\frac{c\pi _{\mu }\sigma
_{\mu }}{\hbar \Omega }u_{+}\left( t,\mathbf{x}\right)  \label{B.23}
\end{equation}%
where 
\begin{equation}
u\left( t,\mathbf{x}\right) =\int u_{-}\left( t,\mathbf{x}\right) dt
\label{B.24}
\end{equation}%
and the arbitrary constant $w_{0}$ is included in the indefinite integral.

Substituting expressions (\ref{B.20}), (\ref{B.23}) in (\ref{B.7c}), we
express the 4-current via solutions $u_{+}$ and $u_{-}$ of the Pauli
equation (\ref{B.17a}). The 4-current $j^{k}$ has regular component $j_{%
\mathrm{reg}}^{k}$ and oscillating component $j_{\mathrm{os}}^{k}$, which
oscillates with the high frequency $\Omega $. We obtain for $j^{k}$ 
\begin{equation*}
j^{k}=j_{\mathrm{reg}}^{k}+j_{\mathrm{os}}^{k},\qquad k=0,1,2,3
\end{equation*}%
\begin{equation}
j_{\mathrm{reg}}^{0}=u_{+}^{\ast }u_{+}+u_{-}^{\ast }u_{-}+\frac{c^{2}}{%
\hbar ^{2}}\left( \pi _{\mu }^{\ast }u^{\ast }\right) \sigma _{\mu }\sigma
_{\nu }\pi _{\nu }u+\frac{1}{4m^{2}c^{2}}\left( \pi _{\mu }^{\ast
}u_{+}^{\ast }\right) \sigma _{\mu }\sigma _{\nu }\pi _{\nu }u_{+}
\label{B.31}
\end{equation}%
\begin{equation}
j_{\mathrm{reg}}^{\mu }=-\frac{1}{2m}\left( u_{+}^{\ast }\sigma _{\mu
}\sigma _{\nu }\pi _{\nu }u_{+}-2i\frac{mc^{2}}{\hbar }u_{-}^{\ast }\sigma
_{\mu }\sigma _{\nu }\pi _{\nu }u\right) +\text{c.c.},\qquad \mu =1,2,3
\label{B31a}
\end{equation}%
\begin{equation*}
j_{\mathrm{os}}^{0}=\left( u_{+}^{\ast }u_{-}+u_{-}^{\ast }u_{+}\right) \cos
\left( \Omega t\right) +i\left( u_{-}^{\ast }u_{+}-u_{+}^{\ast }u_{-}\right)
\sin \left( \Omega t\right)
\end{equation*}%
\begin{equation*}
+\frac{1}{2m\hbar }\left( i\pi _{\mu }^{\ast }u^{\ast }\sigma _{\mu }\sigma
_{\nu }\pi _{\nu }u_{+}-i\pi _{\mu }^{\ast }u_{+}^{\ast }\sigma _{\mu
}\sigma _{\nu }\pi _{\nu }u\right) \cos \left( \Omega t\right)
\end{equation*}%
\begin{equation}
+\frac{1}{2m\hbar }\left( \pi _{\mu }^{\ast }u^{\ast }\sigma _{\mu }\sigma
_{\nu }\pi _{\nu }u_{+}+\pi _{\mu }^{\ast }u_{+}^{\ast }\sigma _{\mu }\sigma
_{\nu }\pi _{\nu }u\right) \sin \left( \Omega t\right)  \label{B.32}
\end{equation}%
\begin{equation*}
j_{\mathrm{os}}^{\mu }=-\frac{1}{2m}\left( u_{-}^{\ast }\sigma _{\mu }\sigma
_{\nu }\pi _{\nu }u_{+}-2i\frac{mc^{2}}{\hbar }u_{+}^{\ast }\sigma _{\mu
}\sigma _{\nu }\pi _{\nu }u\right) \cos \left( \Omega t\right) +\text{c.c}
\end{equation*}%
\begin{equation}
-\frac{1}{2m}\left( -iu_{-}^{\ast }\sigma _{\mu }\sigma _{\nu }\pi _{\nu
}u_{+}+2\frac{mc^{2}}{\hbar }u_{+}^{\ast }\sigma _{\mu }\sigma _{\nu }\pi
_{\nu }u\right) \sin \left( \Omega t\right) +\text{c.c.}  \label{B.32b}
\end{equation}%
where (c.c.) means the expression, which is complex conjugate to the
previous term.

Note that the Dirac particle is charged. It means that the states of the
Dirac particle, containing oscillating charge density $ej^{0}$, or
oscillating current density $e\mathbf{j}$, are unstable with respect to
electromagnetic radiation. As a result of the electromagnetic emanation, the
Dirac particle transits to such a state, where $j^{k}$ does not depend on
time. Such a situation takes place for the states of the electron in the
atom. Only stationary states, where $j^{k}$ does not depend on time, are
stable. The frequency $\Omega =2mc^{2}/\hbar $ is very high, and the
instability is very strong in the sense, that the time of transition to the
stable state is very short.

It follows from the expressions (\ref{B.32}) and (\ref{B.32b}), that if the
state of the Dirac particle contains only low frequencies ($u_{+}\neq 0,\ \
u=0$), or only high frequencies ($u_{+}=0,\ \ u\neq 0$), the oscillating
4-current vanishes ($j_{\mathrm{os}}^{k}=0$, $k=0,1,2,3$). Is it possible
such a situation, that the oscillating 4-current $j_{\mathrm{os}}^{k}$
vanishes at the state, where there are low frequency components and the high
frequency ones together?

To investigate this problem, we use the relation (\ref{B.24}) and rewrite
the expression (\ref{B.32}) for $j_{\mathrm{os}}^{0}$ in the form

\begin{equation*}
j_{\mathrm{os}}^{0}=\left( u_{+}^{\ast }\partial _{0}u-\frac{i}{2m\hbar }%
\left( \pi _{\mu }^{\ast }u_{+}^{\ast }\right) \sigma _{\mu }\sigma _{\nu
}\pi _{\nu }u\right) \cos \left( \Omega t\right) +\text{c.c.}
\end{equation*}%
\begin{equation}
+\left( i\left( \partial _{0}u^{\ast }\right) u_{+}+\frac{1}{2m\hbar }\left(
\pi _{\nu }^{\ast }u^{\ast }\right) \sigma _{\nu }\sigma _{\mu }\pi _{\mu
}u_{+}\right) \sin \left( \Omega t\right) +\text{c.c.}  \label{B.32a}
\end{equation}%
The condition of vanishing $j_{\mathrm{os}}^{0}$ has the form%
\begin{equation}
u_{+}^{\ast }\partial _{0}u-\frac{i}{2m\hbar }\left( \pi _{\mu }^{\ast
}u_{+}^{\ast }\right) \sigma _{\mu }\sigma _{\nu }\pi _{\nu }u=0
\label{B.25}
\end{equation}%
We consider the case, when the electromagnetic field does not depend on $t$,
and hence operator $\partial _{0}$ commutes with operators $\pi _{\mu }$,
defined by (\ref{B.2}). Then according to (\ref{B.24}) and (\ref{B.17a}) the
function $u$ satisfies the equation 
\begin{equation}
i\hbar \partial _{0}u=\hat{H}_{\mathrm{P}}\left( -m\right) u+w_{0}=-\frac{1}{%
2m}\sigma _{\mu }\sigma _{\nu }\pi _{\mu }\pi _{\nu }u+w_{0}  \label{B.26a}
\end{equation}%
where $w_{0}$ is an arbitrary complex two-component constant. We obtain 
\begin{equation}
-\frac{1}{2m}u_{+}^{\ast }\sigma _{\mu }\sigma _{\nu }\pi _{\mu }\pi _{\nu
}u+\frac{1}{2m}\left( \pi _{\mu }^{\ast }u_{+}^{\ast }\right) \sigma _{\mu
}\sigma _{\nu }\pi _{\nu }u+u_{+}^{\ast }w_{0}=0  \label{B.37a}
\end{equation}%
Condition of vanishing $\mathbf{j}_{\mathrm{os}}$ has the form%
\begin{equation}
\pi _{\nu }^{\ast }u_{+}^{\ast }\sigma _{\mu }\sigma _{\nu }u_{-}-2i\frac{%
mc^{2}}{\hbar }u_{+}^{\ast }\sigma _{\mu }\sigma _{\nu }\pi _{\nu
}u=0,\qquad \mu =1,2,3  \label{B.41}
\end{equation}

\section{Plane waves of nonrelativistic Dirac particle}

We describe plane waves of the Dirac particle in terms of the four-component
wave function $\Psi $, defined by the relation%
\begin{equation}
\Psi =\left( 
\begin{array}{c}
\psi _{1} \\ 
\psi _{2}%
\end{array}%
\right) =\frac{\exp \left( -\frac{i}{2}\Omega t\right) }{\sqrt{2}}\left( 
\begin{array}{cc}
I & I \\ 
I & -I%
\end{array}%
\right) \psi  \label{b5.1}
\end{equation}%
where $\psi $ is the dynamical variable of the action (\ref{b1.1}), $\psi
_{1}$, $\psi _{2}$ are dynamical variables of the action (\ref{B.5}), and $I$
is the $2\times 2$ identity matrix. We have two kinds of wave function
describing the plane waves: the low frequency wave function $\Psi _{\mathrm{%
lf}}$ and the high frequency one $\Psi _{\mathrm{hf}}$ 
\begin{equation}
\Psi _{\mathrm{lf}}=\left( 
\begin{array}{c}
\psi _{\mathrm{lf}1} \\ 
\psi _{\mathrm{lf}2}%
\end{array}%
\right) =\exp \left( -i\frac{\mathbf{k}^{2}}{2m\hbar }t+\frac{i\mathbf{kx}}{%
\hbar }\right) \left( 
\begin{array}{c}
\chi \\ 
\frac{\sigma _{\mu }k_{\mu }}{2mc}\chi%
\end{array}%
\right)  \label{b5.2}
\end{equation}%
\begin{equation}
\Psi _{\mathrm{hf}}=\left( 
\begin{array}{c}
\psi _{\mathrm{hf}1} \\ 
\psi _{\mathrm{hf}2}%
\end{array}%
\right) =\exp \left( 2i\frac{mc^{2}t}{\hbar }+i\frac{\mathbf{k}^{2}}{2m\hbar 
}t+\frac{i\mathbf{kx}}{\hbar }\right) \left( 
\begin{array}{c}
-\frac{\sigma _{\mu }k_{\mu }}{2mc}\eta \\ 
\eta%
\end{array}%
\right)  \label{b5.3}
\end{equation}%
where $\chi $ and $\eta $ are two-component constant quantities, and $%
\mathbf{k}=\left\{ k_{1},k_{2},k_{3}\right\} =$const $\left( \mathbf{k}%
^{2}\ll m^{2}c^{2}\right) $ is the canonical momentum of the plane wave. The
plane waves (\ref{b5.2}) and (\ref{b5.3}) are associated with the
nonrelativistic particle and antiparticle. All wave functions $\Psi _{%
\mathrm{lf}}$, $\Psi _{\mathrm{hf}}$ satisfy the nonrelativistic
approximation of the Dirac equation (\ref{B.6}), (\ref{B.7}).

The quantities (\ref{B.7c}), (\ref{B.7d}), (\ref{B.7e}) have the form%
\begin{eqnarray}
j_{\mathrm{lf}}^{0} &=&\left( 1+\frac{\mathbf{k}^{2}}{4m^{2}c^{2}}\right)
\chi ^{\ast }\chi ,\qquad j_{\mathrm{hf}}^{0}=\left( 1+\frac{\mathbf{k}^{2}}{%
4m^{2}c^{2}}\right) \eta ^{\ast }\eta  \label{b5.5} \\
j_{\mathrm{lf}}^{\mu } &=&\frac{k_{\mu }}{m}j_{\mathrm{lf}}^{0}\qquad j_{%
\mathrm{hf}}^{\mu }=-\frac{k_{\mu }}{m}j_{\mathrm{hf}}^{0},  \label{b5.6}
\end{eqnarray}%
\begin{equation}
T_{\mathrm{lf}0}^{0}=\left( mc^{2}+\frac{\mathbf{k}^{2}}{2m}\right) j_{%
\mathrm{lf}}^{0},\qquad T_{\mathrm{hf}0}^{0}=-\left( mc^{2}+\frac{\mathbf{k}%
^{2}}{2m}\right) j_{\mathrm{hf}}^{0},  \label{b5.7}
\end{equation}%
\begin{equation}
T_{\mathrm{lf}0}^{\mu }=k_{\mu }j_{\mathrm{lf}}^{0},\qquad T_{\mathrm{hf}%
0}^{\mu }=-k_{\mu }j_{\mathrm{hf}}^{0}  \label{b5.8}
\end{equation}

Expressions (\ref{b5.5}) -- (\ref{b5.8}) can be obtained also by
conventional method, i.e. as a nonrelativistic approximation of exact
solutions of the Dirac equation in the form of plane waves. The quantities
with index 'lf' are obtained from solution for the positive value of the
temporal component $k_{0}=\left\vert \sqrt{m^{2}c^{2}+\mathbf{k}^{2}}%
\right\vert $ of the canonical momentum, whereas the quantities with index
'hf' are obtained from solution for the negative value of the temporal
component $k_{0}=-\left\vert \sqrt{m^{2}c^{2}+\mathbf{k}^{2}}\right\vert $
of the canonical momentum.

Let us return to investigation of stability conditions (\ref{B.25}), (\ref%
{B.41}). We consider only the case, when $A_{k}=0,\ \ \ \ \pi _{\mu }=i\hbar
\partial _{\mu }$. The low frequency plane wave is associated with the
nonvanishing quantity $u_{+}$, and the high frequency plane wave is
associated with the nonvanishing quantity $u_{-}$ or with the quantity (\ref%
{B.24}). For the plane waves the quantities $u_{+}$, $u$ have the form 
\begin{equation}
u_{+}=\exp \left( i\frac{p_{\alpha }x^{\alpha }}{\hbar }-i\frac{\mathbf{p}%
^{2}}{2m\hbar }t\right) \chi _{+},\qquad u=-i\frac{2m\hbar }{\mathbf{k}^{2}}%
\exp \left( i\frac{k_{\alpha }x^{\alpha }}{\hbar }+i\frac{\mathbf{k}^{2}}{%
2m\hbar }t\right) \chi _{-}  \label{B.38}
\end{equation}%
\begin{equation}
\mathbf{k}^{2},\mathbf{p}^{2}\ll m^{2}c^{2},\qquad \chi _{+},\chi _{-}=\text{%
const}  \label{B.38a}
\end{equation}
where $\chi _{+}$, $\chi _{-}$ are two-component constant quantities.
Substituting (\ref{B.38}) in the constraints (\ref{B.37a}), (\ref{B.41}), we
obtain 
\begin{subequations}
\begin{eqnarray}
\chi _{+}^{\ast }\chi _{-}-\frac{1}{\mathbf{k}^{2}}p_{\mu }k_{\nu }\chi
_{+}^{\ast }\sigma _{\mu }\sigma _{\nu }\chi _{-} &=&0  \label{B.39} \\
\chi _{+}^{\ast }\sigma _{\mu }\sigma _{\nu }\chi _{-}\left( -p_{\nu }+\frac{%
4m^{2}c^{2}}{\mathbf{k}^{2}}k_{\nu }\right) &=&0,\qquad \mu =1,2,3
\label{B.40}
\end{eqnarray}

Eliminating $\chi _{+}^{\ast }\sigma _{\mu }\sigma _{\nu }\chi _{-}$ from
equations (\ref{B.39}) and (\ref{B.40}), we obtain 
\end{subequations}
\begin{equation}
\left( -p_{\mu }p_{\mu }+4m^{2}c^{2}\right) \chi _{+}^{\ast }\chi _{-}=0
\label{B.40a}
\end{equation}%
As far as $\left\vert \mathbf{p}\right\vert \ll 2mc$, the bracket in (\ref%
{B.40a}) cannot vanish, and we obtain%
\begin{equation}
\chi _{+}^{\ast }\chi _{-}=0  \label{a5.1}
\end{equation}%
Applying the identity (\ref{B.7f}) to the relation (\ref{B.40}) and taking
into account (\ref{a5.1}), we obtain%
\begin{equation}
-i\varepsilon _{\mu \nu \alpha }\chi _{+}^{\ast }\sigma _{\alpha }\chi
_{-}\left( -p_{\nu }+\frac{4m^{2}c^{2}}{\mathbf{k}^{2}}k_{\nu }\right)
=0,\qquad \mu =1,2,3  \label{a5.2}
\end{equation}%
As far as the bracket in (\ref{a5.2}) cannot vanish, we obtain from (\ref%
{a5.2}) 
\begin{equation}
\chi _{+}^{\ast }\sigma _{\mu }\chi _{-}=q_{\mu },\qquad q_{\mu }=A\left(
-p_{\mu }+\frac{4m^{2}c^{2}}{\mathbf{k}^{2}}k_{\mu }\right) ,\qquad \mu
=1,2,3  \label{a5.3}
\end{equation}%
where $A$ is some complex number.

Let us represent the quantities $\chi _{+}$, $\chi _{-}$ in the form%
\begin{equation}
\chi _{+}=\left( 
\begin{array}{c}
a_{1} \\ 
a_{2}%
\end{array}%
\right) ,\qquad \chi _{-}=\left( 
\begin{array}{c}
b_{1} \\ 
b_{2}%
\end{array}%
\right)  \label{a5.4}
\end{equation}%
Then relations (\ref{a5.1}) and (\ref{a5.2}) take the form%
\begin{equation*}
a_{1}^{\ast }b_{1}+a_{2}^{\ast }b_{2}=0,\qquad a_{1}^{\ast
}b_{2}+a_{2}^{\ast }b_{1}=q_{1},
\end{equation*}

\begin{equation*}
-ia_{1}^{\ast }b_{2}+ia_{2}^{\ast }b_{1}=q_{2},\qquad a_{1}^{\ast
}b_{1}-a_{2}^{\ast }b_{2}=q_{3}
\end{equation*}%
These equations are transformed to the form%
\begin{equation}
a_{1}^{\ast }b_{1}=\frac{q_{3}}{2},\qquad a_{2}^{\ast }b_{1}=\frac{%
q_{1}-iq_{2}}{2},\qquad a_{1}^{\ast }b_{2}=\frac{q_{1}+iq_{2}}{2},\qquad
a_{2}^{\ast }b_{2}=-\frac{q_{3}}{2}  \label{a5.5}
\end{equation}

It follows from the first two relations (\ref{a5.5}) and from the last two
relations (\ref{a5.5})%
\begin{equation}
\frac{a_{1}^{\ast }}{a_{2}^{\ast }}=\frac{q_{3}}{\left( q_{1}-iq_{2}\right) }%
,\qquad \frac{a_{1}^{\ast }}{a_{2}^{\ast }}=-\frac{q_{1}+iq_{2}}{q_{3}}
\label{a5.6}
\end{equation}%
Two relations (\ref{a5.6}) are compatible, only if%
\begin{equation}
q_{3}^{2}+q_{1}^{2}+q_{2}^{2}=A^{2}\left( -p_{\mu }+\frac{4m^{2}c^{2}}{%
\mathbf{k}^{2}}k_{\mu }\right) \left( -p_{\mu }+\frac{4m^{2}c^{2}}{\mathbf{k}%
^{2}}k_{\mu }\right) =0  \label{a5.7}
\end{equation}%
As far as the brackets in (\ref{a5.7}) cannot vanish, the relation (\ref%
{a5.7}) can be satisfied, only if $A=0$ and $q_{\mu }=0$, $\mu =1,2,3$. Then
we obtain from (\ref{a5.5})%
\begin{equation}
a_{1}^{\ast }b_{1}=0,\qquad a_{2}^{\ast }b_{2}=0,\qquad a_{1}^{\ast
}b_{2}=0,\qquad a_{2}^{\ast }b_{1}=0  \label{a5.8}
\end{equation}

If $a_{1}\neq 0\vee a_{2}\neq 0$, the equations (\ref{a5.8}) can be
satisfied only if $b_{1}=0\wedge b_{2}=0$. If $b_{1}\neq 0\vee b_{2}\neq 0$,
the equations (\ref{a5.8}) can be satisfied only if $a_{1}=0\wedge a_{2}=0$.
It means that any linear combination of the low frequency solution and of
the high frequency solution is unstable with respect to electromagnetic
radiation.

Stable superposition of $\Psi _{\mathrm{lf}}$ and of $\Psi _{\mathrm{hf}}$
is impossible. In the stable states $\Psi _{\mathrm{lf}}$ and $\Psi _{%
\mathrm{hf}}$ can be considered as states of different dynamic systems. In
other words, in the stable states of the Dirac particle the superselection
rule takes place.

\section{Classical approximation of the nonrelativistic Dirac particle}

To make the dynamic disquantization, we need to introduce hydrodynamic
variables, where the current $j^{k}$ were the dependent variable instead of $%
\psi $. Transforming the action (\ref{b1.1}), we use the mathematical
technique \cite{S30,S51}, where the wave function $\psi $ is considered to
be a function of hypercomplex numbers $\gamma $ and coordinates $x$. In this
case the physical quantities are obtained by means of a convolution of
expressions $\psi ^{\ast }O\psi $ with the zero divisor. This technique
allows one to work without fixing the $\gamma $-matrices representation.

Using designations 
\begin{equation}
\gamma _{5}=c\gamma ^{0123}\equiv c\gamma ^{0}\gamma ^{1}\gamma ^{2}\gamma
^{3},  \label{f1.9}
\end{equation}
\begin{equation}
\mathbf{\sigma }=\{\sigma _{1},\sigma _{2},\sigma _{3},\}=\{-i\gamma
^{2}\gamma ^{3},-i\gamma ^{3}\gamma ^{1},-i\gamma ^{1}\gamma ^{2}\}
\label{f1.10}
\end{equation}
we make the change of variables 
\begin{equation}
\psi =Ae^{i\varphi +{\frac{1}{2}}\gamma _{5}\kappa }\exp \left( -\frac{i}{2}%
\gamma _{5}\mathbf{\sigma \eta }\right) \exp \left( {\frac{i\pi }{2}}\mathbf{%
\sigma n}\right) \Pi  \label{f1.11}
\end{equation}
\begin{equation}
\psi ^{\ast }=A\Pi \exp \left( -{\frac{i\pi }{2}}\mathbf{\sigma n}\right)
\exp \left( -\frac{i}{2}\gamma _{5}\mathbf{\sigma \eta }\right) e^{-i\varphi
-{\frac{1}{2}}\gamma _{5}\kappa }  \label{f1.12}
\end{equation}
where (*) means the Hermitian conjugation, and the quantity 
\begin{equation}
\Pi ={\frac{1}{4}}(1+c\gamma ^{0})(1+\mathbf{z\sigma }),\qquad \mathbf{z}%
=\{z^{\alpha }\}=\text{const},\qquad \alpha =1,2,3;\qquad \mathbf{z}^{2}=1
\label{f1.13}
\end{equation}
is the zero divisor (projector). The quantities $A$, $\kappa $, $\varphi $, $%
\mathbf{\eta }=\{\eta ^{\alpha }\}$, $\mathbf{n}=\{n^{\alpha }\}$, $\alpha
=1,2,3,\;$ $\mathbf{n}^{2}=1$ are eight real parameters, determining the
wave function $\psi .$ These parameters may be considered as new dependent
variables, describing the state of dynamic system $\mathcal{S}_{\mathrm{D}}$%
. The quantity $\varphi $ is a scalar, and $\kappa $ is a pseudoscalar. Six
remaining variables $A,$ $\mathbf{\eta }=\{\eta ^{\alpha }\}$, $\mathbf{n}%
=\{n^{\alpha }\}$, $\alpha =1,2,3,\;$ $\mathbf{n}^{2}=1$ can be expressed
through the flux 4-vector $j^{l}=\bar{\psi}\gamma ^{l}\psi $ and spin
4-pseudovector 
\begin{equation}
S^{l}=i\bar{\psi}\gamma _{5}\gamma ^{l}\psi ,\qquad l=0,1,2,3  \label{f1.13a}
\end{equation}
Because of two identities 
\begin{equation}
S^{l}S_{l}\equiv -j^{l}j_{l},\qquad j^{l}S_{l}\equiv 0.  \label{f1.14}
\end{equation}
there are only six independent components among eight components of
quantities $j^{l}$, and $S^{l}$.

Mathematical details of the dependent variables transformation can be found
in \cite{R2004}, where the action is calculated for the case $c=1$ and
vanishing electromagnetic field $A_{l}=0$. As a result we have the following
form of the action, written in the hydrodynamical form 
\begin{equation}
\mathcal{S}_{\mathrm{D}}:\qquad \mathcal{A}_{D}[j,\varphi ,\kappa ,\mathbf{%
\xi }]=\int \mathcal{L}d^{4}x,\qquad \mathcal{L}=\mathcal{L}_{\mathrm{cl}}+%
\mathcal{L}_{\mathrm{q1}}+\mathcal{L}_{\mathrm{q2}}  \label{c4.15}
\end{equation}
\begin{equation}
\mathcal{L}_{\mathrm{cl}}=-mc\rho -\hbar j^{l}\partial _{l}\varphi -\frac{e}{%
c}A_{l}j^{l}-\frac{\hbar j^{l}}{2\left( 1+\mathbf{\xi z}\right) }\varepsilon
_{\alpha \beta \gamma }\xi ^{\alpha }\partial _{l}\xi ^{\beta }z^{\gamma
},\qquad \rho \equiv \sqrt{j^{l}j_{l}}  \label{c4.16}
\end{equation}
\begin{equation}
\mathcal{L}_{\mathrm{q1}}=2mc\rho \sin ^{2}(\frac{\kappa }{2})-{\frac{\hbar 
}{2}}S^{l}\partial _{l}\kappa ,  \label{c4.17}
\end{equation}
\begin{equation}
\mathcal{L}_{\mathrm{q2}}=\frac{\hbar (\rho +cj^{0})}{2}\varepsilon _{\alpha
\beta \gamma }\partial ^{\alpha }\frac{j^{\beta }}{(\rho +cj^{0})}\xi
^{\gamma }-\frac{\hbar }{2(\rho +cj^{0})}\varepsilon _{\alpha \beta \gamma
}\left( \partial ^{0}j^{\beta }\right) j^{\alpha }\xi ^{\gamma }
\label{c4.18}
\end{equation}
where $\varepsilon _{\alpha \beta \gamma }$ is the Levi-Chivita
3-pseudotensor. The Lagrangian density $\mathcal{L}$ is a function of
4-vector $j^{l}$, scalar $\varphi $, pseudoscalar $\kappa $, and the unit
3-pseudovector $\mathbf{\xi }$, which is connected with the spin
4-pseudovector $S^{l}$ by means of the relations 
\begin{equation}
\xi ^{\alpha }=\rho ^{-1}\left[ S^{\alpha }-\frac{j^{\alpha }S^{0}}{(\rho
+cj^{0})}\right] ,\qquad \alpha =1,2,3;\qquad \rho \equiv \sqrt{j^{l}j_{l}}
\label{f1.15}
\end{equation}
\begin{equation}
S^{0}=\mathbf{j\xi },\qquad S^{\alpha }=\rho \xi ^{\alpha }+\frac{(\mathbf{%
j\xi })j^{\alpha }}{\rho +cj^{0}},\qquad \alpha =1,2,3  \label{f1.16}
\end{equation}

Producing the dynamical disquantization (\ref{b2.5}) in (\ref{c4.15}) - (\ref%
{c4.18}), we obtain 
\begin{equation*}
\mathcal{A}_{\mathrm{Dqu}}[j,\varphi ,\mathbf{\xi }]=\int \left\{ -\kappa
_{0}m\rho -\frac{e}{c}A_{l}j^{l}-\hbar j^{i}\left( \partial _{i}\varphi +%
\frac{\varepsilon _{\alpha \beta \gamma }\xi _{\alpha }\partial _{i}\xi
_{\beta }z_{\gamma }}{2\left( 1+\mathbf{\xi z}\right) }\right) \right.
\end{equation*}%
\begin{equation}
+\left. \frac{\hbar j^{k}}{2(\rho +j_{0})\rho }\varepsilon _{\alpha \beta
\gamma }\left( \partial _{k}j^{\beta }\right) j^{\alpha }\xi _{\gamma
}\right\} d^{4}x  \label{a5.13a}
\end{equation}%
where $\kappa _{0}=\pm 1$ is the solution of the dynamic equation $\delta 
\mathcal{A}_{\mathrm{Dqu}}/\delta \kappa =0$, which does not contain
derivatives, because the last term of (\ref{c4.17}) vanishes after dynamical
disquantization (\ref{b2.5}) in virtue of the second identity (\ref{f1.14}).

We introduce the Lagrangian coordinates $\tau =\{\tau _{0},\mathbf{\tau }%
\}=\{\tau _{i}\left( x\right) \}$, $i=0,1,2,3$ as functions of coordinates $%
x $ in such a way that only coordinate $\tau _{0}$ changes along the
direction $j^{l}$, i.e. 
\begin{equation}
j^{k}\partial _{k}\tau _{\mu }=0,\qquad \mu =1,2,3  \label{b3.1}
\end{equation}
Considering the variables $\tau =\{\tau _{0},\mathbf{\tau }\}$ as
independent variables in (\ref{a5.13a}), we obtain after calculations (See
mathematical details in \cite{R2004}) 
\begin{equation}
\mathcal{A}_{\mathrm{Dqu}}[x,\mathbf{\xi }]=\int \left\{ -\kappa _{0}mc\sqrt{%
\dot{x}^{l}\dot{x}_{l}}-\frac{e}{c}A_{l}\dot{x}^{l}+\hbar {\frac{(\dot{%
\mathbf{\xi }}\times \mathbf{\xi })\mathbf{z}}{2(1+\mathbf{\xi z})}}+\hbar 
\frac{(\dot{\mathbf{x}}\times \ddot{\mathbf{x}})\mathbf{\xi }}{2\sqrt{\dot{x}%
^{s}\dot{x}_{s}}(\sqrt{\dot{x}^{s}\dot{x}_{s}}+c\dot{x}^{0})}\right\}
d^{4}\tau  \label{a5.18}
\end{equation}
where period means the total derivative $\dot{x}^{s}\equiv dx^{s}/d\tau _{0}$%
. The quantities \ $x=\left\{ x^{0},\mathbf{x}\right\} =\{x^{i}\}$, $%
\;i=0,1,2,3$, and $\mathbf{\xi }=\{\xi _{\alpha }\}$, $\alpha =1,2,3$ are
considered to be functions of the Lagrangian coordinates $\tau _{0}$, $%
\mathbf{\tau }=\{\tau _{1},\tau _{2},\tau _{3}\}$. Here and in what follows
the symbol $\times $ means the vector product of two 3-vectors. The quantity$%
\;\mathbf{z}$ is the constant unit 3-vector (\ref{f1.13}). In fact,
variables $x$ depend on $\mathbf{\tau }$ as on parameters, because the
action (\ref{a5.18}) does not contain derivatives with respect to $\tau
_{\alpha }$, \ $\alpha =1,2,3$. Lagrangian density of the action (\ref{a5.18}%
) does not contain independent variables $\tau $ explicitly. Hence, it may
be written in the form 
\begin{equation}
\mathcal{A}_{\mathrm{Dqu}}[x,\mathbf{\xi }]=\int \mathcal{A}_{\mathrm{Dcl}%
}[x,\mathbf{\xi }]d\mathbf{\tau ,\qquad }d\mathbf{\tau }=d\tau _{1}d\tau
_{2}d\tau _{3}  \label{b3.8}
\end{equation}
where 
\begin{equation}
\mathcal{A}_{\mathrm{Dcl}}[x,\mathbf{\xi }]=\int \left\{ -\kappa _{0}mc\sqrt{%
\dot{x}^{i}\dot{x}_{i}}-\frac{e}{c}A_{l}\dot{x}^{l}+\hbar {\frac{(\dot{%
\mathbf{\xi }}\times \mathbf{\xi })\mathbf{z}}{2(1+\mathbf{\xi z})}}+\hbar 
\frac{(\dot{\mathbf{x}}\times \ddot{\mathbf{x}})\mathbf{\xi }}{2\sqrt{\dot{x}%
^{s}\dot{x}_{s}}(\sqrt{\dot{x}^{s}\dot{x}_{s}}+c\dot{x}^{0})}\right\} d\tau
_{0}  \label{b3.9}
\end{equation}

It is easy to see that the action (\ref{b3.9}) is invariant with respect to
transformation $\tau _{0}\rightarrow \tilde{\tau}_{0}=F\left( \tau
_{0}\right) $, where $F$ is an arbitrary monotone function. This invariance
admits one to choose the variable $t=x^{0}$ as a parameter $\tau _{0}$. In
this case we obtain instead of (\ref{b3.9}) 
\begin{eqnarray}
\mathcal{A}_{\mathrm{Dcl}}[\mathbf{x},\mathbf{\xi }] &=&\int \left\{ -\kappa
_{0}mc^{2}\sqrt{1-\frac{\mathbf{\dot{x}}^{2}}{c^{2}}}-\frac{e}{c}A_{0}-\frac{%
e}{c}\mathbf{A\dot{x}}+\hbar {\frac{(\dot{\mathbf{\xi }}\times \mathbf{\xi })%
\mathbf{z}}{2(1+\mathbf{\xi z})}}\right.  \notag \\
&&\left. +\frac{\hbar }{2c^{2}}(\dot{\mathbf{x}}\times \ddot{\mathbf{x}})%
\mathbf{\xi }\left( 1-\frac{\mathbf{\dot{x}}^{2}}{c^{2}}+\sqrt{1-\frac{%
\mathbf{\dot{x}}^{2}}{c^{2}}}\right) ^{-1}\right\} dt  \label{b3.10}
\end{eqnarray}

In the nonrelativistic approximation, when $c\rightarrow \infty $, the
coefficient $\frac{\hbar }{2c^{2}}$ before the last term tends to zero.
Nevertheless, we may not omit the last term in the action (\ref{b3.10}),
because the last term contains the highest derivative. In the dynamic
equations this term generates the term with the small parameter before the
highest derivative. Such a term may not be omitted, because it is of the
same order as the other terms. This term may generate oscillations with the
frequency of the order $\Omega =2mc^{2}/\hbar $, and $\Omega \hbar
/mc^{2}\approx 1$.

In the nonrelativistic approximation the action (\ref{b3.10}) turns into 
\begin{equation}
\mathcal{A}_{\mathrm{Dcl}}[\mathbf{x},\mathbf{\xi }]=\int \left\{ \frac{1}{2}%
\kappa _{0}m\mathbf{\dot{x}}^{2}+\frac{\hbar }{4c^{2}}(\dot{\mathbf{x}}%
\times \ddot{\mathbf{x}})\mathbf{\xi }-\frac{e}{c}\left( A_{0}+\mathbf{A\dot{%
x}}\right) +\hbar {\frac{(\dot{\mathbf{\xi }}\times \mathbf{\xi })\mathbf{z}%
}{2(1+\mathbf{\xi z})}}\right\} dt  \label{b3.11}
\end{equation}
where the first term $-\kappa _{0}mc^{2}$ is omitted, because it gives no
contribution in the dynamic equations. Two first terms in (\ref{b3.11})
describe dynamics and structure of the classical Dirac particle. The third
term describes interaction with the electromagnetic field.

\section{Solution of dynamic equation for the classical nonrelativistic
Dirac particle}

Dynamic equations for the classical nonrelativistic Dirac particle $\mathcal{%
S}_{\mathrm{nDcl}}$ generated by the action (\ref{b3.11}) have the form 
\begin{equation}
-\kappa _{0}m\mathbf{\ddot{x}-}e\mathbf{E-}\frac{e}{c}\left( \mathbf{\dot{x}}%
\times \mathbf{H}\right) \mathbf{+}\hbar \frac{d}{dt}\frac{(\mathbf{\xi }%
\times \ddot{\mathbf{x}})}{2c^{2}}-\hbar \frac{d}{dt}\frac{(\dot{\mathbf{x}}%
\times \mathbf{\dot{\xi}})}{4c^{2}}=0  \label{f2.16}
\end{equation}
where 
\begin{equation}
\mathbf{E}=-\frac{1}{c}\frac{\partial \mathbf{A}}{\partial t}+\frac{1}{c}%
\mathbf{\nabla }A_{0},\qquad \mathbf{H}=\mathbf{\nabla }\times \mathbf{%
A,\qquad }A_{k}=\left\{ A_{0},\mathbf{A}\right\}  \label{g2.18}
\end{equation}
\begin{equation}
\mathbf{\xi }\times \left( -\hbar {\frac{(\dot{\mathbf{\xi }}\times \mathbf{z%
})}{2(1+\mathbf{\xi z})}+}\hbar \frac{d}{dt}{\frac{(\mathbf{z}\times \mathbf{%
\xi })}{2(1+\mathbf{\xi }_{\mu }\mathbf{z})}-}\hbar {\frac{\left( (\dot{%
\mathbf{\xi }}\times \mathbf{\xi })\mathbf{z}\right) }{2(1+\mathbf{\xi z}%
)^{2}}}\mathbf{z}+\hbar \frac{(\dot{\mathbf{x}}\times \ddot{\mathbf{x}})}{%
4c^{2}}\right) =0  \label{f2.17}
\end{equation}
Vector product in (\ref{f2.17}) is a corollary of the constraint $\mathbf{%
\xi }^{2}=1$.

After simplification the dynamic equation (\ref{f2.17}) is reduced to the
form (See Appendix B of \cite{R2004}) 
\begin{equation}
{\dot{\mathbf{\xi }}}=\frac{\mathbf{\xi }\times (\dot{\mathbf{x}}\times 
\ddot{\mathbf{x}})}{2c^{2}}  \label{f2.18}
\end{equation}
This equation describes rotation of the unit vector $\mathbf{\xi }$ with the
angular frequency $\omega =c^{-2}\left( \dot{\mathbf{x}}\times \ddot{\mathbf{%
x}}\right) /2$. In general, we may not neglect the rhs of (\ref{f2.18}) in
the nonrelativistic approximation, if $\mathbf{\dot{x}}$ oscillates with the
frequency of the order of $mc^{2}/\hbar $. Solving dynamic equations (\ref%
{f2.16}), (\ref{f2.17}) we shall see that such frequencies are possible.

When the electromagnetic field is absent ($\mathbf{E}=0$ and $\mathbf{H}=0$%
), the equation (\ref{f2.16}) is reduced to the form 
\begin{equation}
-\kappa _{0}m\mathbf{\ddot{x}+}\hbar \frac{d}{dt}\frac{(\mathbf{\xi }\times 
\ddot{\mathbf{x}})}{2c^{2}}-\hbar \frac{d}{dt}\frac{(\dot{\mathbf{x}}\times 
\mathbf{\dot{\xi}})}{4c^{2}}=0  \label{f2.19}
\end{equation}
\begin{equation}
-\kappa _{0}m\mathbf{\dot{x}+}\hbar \frac{(\mathbf{\xi }\times \ddot{\mathbf{%
x}})}{2c^{2}}-\hbar \frac{(\dot{\mathbf{x}}\times \mathbf{\dot{\xi}})}{4c^{2}%
}=-\mathbf{p}=\text{const}  \label{f2.20}
\end{equation}
The equation (\ref{f2.20}) can be solved exactly, if $\dot{\mathbf{x}}\times 
\ddot{\mathbf{x}}=a\mathbf{\xi }$, where $a$ is an arbitrary quantity. Then
according to (\ref{f2.18}) 
\begin{equation}
\mathbf{\xi }=\text{const}  \label{f2.21}
\end{equation}
The equation (\ref{f2.20}) turns into 
\begin{equation}
-\kappa _{0}m\mathbf{\dot{x}+}\hbar \frac{(\mathbf{\xi }\times \ddot{\mathbf{%
x}})}{2c^{2}}=-\mathbf{p}  \label{f2.22}
\end{equation}
The general solution of (\ref{f2.22}) has the form 
\begin{equation}
\mathbf{\dot{x}=}\frac{\mathbf{p}}{\kappa _{0}m}+\mathbf{V}\cos \left(
\omega t+\phi \right) +\mathbf{\xi }\times \mathbf{V}\sin \left( \omega
t+\phi \right) ,\qquad \left\vert \mathbf{V}\right\vert \ll c  \label{f2.23}
\end{equation}
where $\phi $ is an arbitrary constant. The quantities $\mathbf{V}$ and $%
\mathbf{p}$ are the constant vectors satisfying the constraints 
\begin{equation}
\mathbf{V\xi }=0,\qquad \mathbf{\xi }^{2}=1  \label{f2.24}
\end{equation}
and the frequency $\omega $ is determined by the relation 
\begin{equation}
\omega =-\kappa _{0}\frac{2mc^{2}}{\hbar },\qquad \kappa _{0}=\pm 1
\label{f2.25}
\end{equation}
After integration of (\ref{f2.23}) we obtain 
\begin{equation}
\mathbf{x=X+}\frac{\mathbf{p}t}{\kappa _{0}m}-\mathbf{V}\frac{\kappa
_{0}\hbar }{2mc^{2}}\sin \left( \omega t+\phi \right) +\mathbf{\xi }\times 
\mathbf{V}\frac{\kappa _{0}\hbar }{2mc^{2}}\cos \left( \omega t+\phi \right)
,\qquad \mathbf{X}=\text{const}  \label{f2.26}
\end{equation}

Thus, the world line of the free nonrelativistic classical Dirac particle $%
\mathcal{S}_{\mathrm{nDcl}}$ is a helix. According to the condition $%
\left\vert \mathbf{V}\right\vert \ll c$ the radius $r$ of the helix is much
less, than the Compton wave length $\lambda _{\mathrm{C}}=\hbar /mc$. This
result agrees with the result of investigation of the relativistic classical
Dirac particle $\mathcal{S}_{\mathrm{Dcl}}$ \cite{R2004}, where the world
line is also a helix, but without the constraint $\left\vert \mathbf{V}%
\right\vert \ll c$. In the limit $c\rightarrow \infty $ the oscillating
terms vanish in expression (\ref{f2.26}) for $\mathbf{x}$. However, they are
not vanish in the expression (\ref{f2.23}) for $\mathbf{\dot{x}}$. In the
limit $c\rightarrow \infty $ the helix turns into straight line, but the
velocity of circular motion does not vanish.

The angular momentum generated by the solution (\ref{f2.26}), (\ref{f2.23})
has the form 
\begin{equation}
\mathbf{M}=m\left( \mathbf{x}\times \mathbf{\dot{x}}\right) =\frac{\mathbf{%
\xi V}^{2}m}{\omega }+\mathbf{M}_{\mathrm{os}}  \label{f2.26b}
\end{equation}
where $\mathbf{M}_{\mathrm{os}}$ is the oscillating part of the angular
momentum. Averaging over the time, the mean value $\left\langle \mathbf{M}_{%
\mathrm{os}}\right\rangle $ of $\mathbf{M}_{\mathrm{os}}$ vanishes. Then 
\begin{equation}
\left\langle \mathbf{M}\right\rangle =m\left( \mathbf{x}\times \mathbf{\dot{x%
}}\right) =-\kappa _{0}\frac{\hbar }{2}\mathbf{\xi }\frac{\mathbf{V}^{2}}{%
c^{2}},\qquad \kappa _{0}=\pm 1  \label{f2.26c}
\end{equation}
This result is applicable only in the nonrelativistic case, when $V^{2}\ll
c^{2}$. The mean angular momentum $\left\langle \mathbf{M}\right\rangle $ is
directed along the unit vector $\mathbf{\xi }$. Its module is equal to $%
\hbar /2$, provided $V=c$.

In the general relativistic case the velocity amplitude $\left\vert \mathbf{V%
}\right\vert $ and the frequency $\omega $ are connected between themselves,
and the solution (\ref{f2.26}) turns \cite{R2004} into the relation 
\begin{equation}
\mathbf{x=X}+\frac{\mathbf{V}}{\Omega _{\mathrm{Dcl}}}\sin \left( \Omega _{%
\mathrm{Dcl}}t+\phi \right) -\frac{\mathbf{\xi }\times \mathbf{V}}{\Omega _{%
\mathrm{Dcl}}}\cos \left( \Omega _{\mathrm{Dcl}}t+\phi \right) ,\qquad 
\mathbf{X}=\text{const}  \label{f2.27}
\end{equation}
where 
\begin{equation}
\mathbf{V}=c\frac{\sqrt{\gamma ^{2}-1}}{\gamma }\mathbf{n},\qquad \Omega _{%
\mathrm{Dcl}}=-\kappa _{0}\frac{2mc^{2}}{\hbar \gamma ^{2}},\qquad \mathbf{n}%
^{2}=1,\qquad \mathbf{n\xi }=0  \label{f2.28}
\end{equation}
and $\gamma \geq 1$ is an arbitrary constant (Lorentz-factor of rotation).
The regular momentum $\mathbf{p}=0$, because only in this case one succeeded
to solve exactly the relativistic dynamic equations. At $\gamma \rightarrow
1 $ the relation (\ref{f2.27}) turns into (\ref{f2.26}) with $\mathbf{p}=0$.

In the relativistic case the mean magnetic moment has the form 
\begin{equation}
\left\langle \mathbf{M}_{\mathrm{Dcl}}\right\rangle =m\left( \mathbf{x}%
\times \mathbf{\dot{x}}\right) =-\kappa _{0}V^{2}\frac{\hbar }{2c^{2}}\gamma
^{2}\mathbf{\xi }=-\kappa _{0}\left( \gamma ^{2}-1\right) \frac{\hbar }{2}%
\mathbf{\xi }  \label{f2.29}
\end{equation}

In the Dirac dynamic system $\mathcal{S}_{\mathrm{D}}$ the internal degrees
of freedom are described nonrelativistically \cite{R2004a}. This defect can
be corrected \cite{R2004a}. After such a correction the classical Dirac
particle $\mathcal{S}_{\mathrm{Dcl}}$ turns into the modified classical
Dirac particle $\mathcal{S}_{\mathrm{mDcl}}$. In this case we have instead
of (\ref{f2.27}) - (\ref{f2.29}) 
\begin{equation}
\mathbf{x=X}+\frac{\mathbf{V}}{\Omega _{\mathrm{mDcl}}}\sin \left( \Omega _{%
\mathrm{mDcl}}t+\phi \right) -\frac{\mathbf{\xi }\times \mathbf{V}}{\Omega _{%
\mathrm{mDcl}}}\cos \left( \Omega _{\mathrm{mDcl}}t+\phi \right) ,\qquad 
\mathbf{X}=\text{const}  \label{f2.30}
\end{equation}
\begin{equation}
\mathbf{V}=c\frac{\sqrt{\gamma ^{2}-1}}{\gamma }\mathbf{n},\qquad \Omega _{%
\mathrm{mDcl}}=-\kappa _{0}\frac{2mc^{2}}{\hbar \left( 2\gamma ^{2}-1\right)
^{2}},\qquad \mathbf{n}^{2}=1,\qquad \mathbf{n\xi }=0  \label{f2.31}
\end{equation}

\begin{equation}
\left\langle \mathbf{M}_{\mathrm{mDcl}}\right\rangle =m\left( \mathbf{x}%
\times \mathbf{\dot{x}}\right) =-\kappa _{0}\frac{\left( \gamma
^{2}-1\right) }{\gamma ^{2}}\left( 2\gamma ^{2}-1\right) ^{2}\frac{\hbar }{2}%
\mathbf{\xi }  \label{f2.32}
\end{equation}%
In the nonrelativistic case, when $\gamma -1\ll 1$, the results (\ref{f2.32}%
), (\ref{f2.29}) coincide with (\ref{f2.26c}) and between themselves.

Conventionally, the nonrelativistic approximation is obtained by other
method (See, for instance, \cite{D58}, sec. 70). One derives dynamic
equations for nonrelativistic Dirac particle, and thereafter one transits to
the classical approximation (dynamic disquantization). As a result one
obtains the relation (\ref{f2.26}) without two last terms, i.e. $\mathbf{V}%
\equiv 0$, and one obtains a straight line instead of a helix. The loss of
the two last terms in (\ref{f2.26}) takes place at the stage of transition
from the Dirac equation to the Pauli equation. Formally the loss of two last
terms in (\ref{f2.26}) is justified by the fact that these terms are
proportional to $c^{-2}$ and small in the limit $c\rightarrow \infty $.

We are to remark here that the nonrelativistic approximation is described by
the inequality $\left\vert \mathbf{\dot{x}}\right\vert \ll c$. It concerns
only velocities and does not impose any constraints on the position $\mathbf{%
x}$. As to the velocity, all terms in the relation (\ref{f2.23}) for the
velocity are of the same order, and we may not neglect the two last terms in
(\ref{f2.23}). The additional terms give a very small contribution to the
particle position, but they introduce additional degrees of freedom, which
are rather rigid . These degrees of freedom cannot be excited at the low
energies, characteristic for the atomic spectra. The characteristic energy,
connected with these degrees of freedom is of the order $\hbar \omega
=2mc^{2}$. In other words, it is a characteristic threshold energy of the
pair production.

\section{Relativistic corrections to nonrelativistic \newline
classical Dirac particle}

Note that the action (\ref{b3.11}) for the nonrelativistic classical Dirac
particle as well as the dynamic equation (\ref{f2.16}) does not contain the
term, describing interaction of the magnetic moment with the magnetic field,
what is characteristic for the nonrelativistic Pauli equation. This
interaction may be obtained, if we take the high frequency solution (\ref%
{f2.23}) and average the action (\ref{b3.11}) over the frequency $\omega $,
determined by the relation (\ref{f2.25}). In reality, it is necessary to
average only the term $-\frac{e}{c}\mathbf{A}\left( t,\mathbf{x}\right) 
\mathbf{\dot{x}}$. We obtain 
\begin{equation}
\left\langle -\frac{e}{c}\mathbf{A}\left( t,\mathbf{x}\right) \mathbf{\dot{x}%
}\right\rangle =-\frac{e}{c}\left\langle \mathbf{A}\left( t,\mathbf{x}%
\right) \delta \mathbf{\dot{x}}+\delta x^{\mu }\partial _{\mu }\mathbf{A}%
\left( t,\mathbf{x}\right) \delta \mathbf{\dot{x}}\right\rangle  \label{c7.1}
\end{equation}
where $\delta \mathbf{x}$ and $\delta \mathbf{\dot{x}}$ are determined by
the relations (\ref{f2.26}) and (\ref{f2.23}) 
\begin{equation}
\delta \mathbf{x=}-\frac{\mathbf{V}}{\omega }\sin \left( \omega t\right) +%
\mathbf{\xi }\times \frac{\mathbf{V}}{\omega }\cos \left( \omega t\right)
,\qquad \omega =-\kappa _{0}\frac{2mc^{2}}{\hbar },\qquad \kappa _{0}=\pm 1
\label{c7.2}
\end{equation}
\begin{equation}
\delta \mathbf{\dot{x}=}\frac{\mathbf{p}}{\kappa _{0}m}+\mathbf{V}\cos
\left( \omega t\right) +\mathbf{\xi }\times \mathbf{V}\sin \left( \omega
t\right) ,\qquad \mathbf{V\xi }=0,\qquad \mathbf{\xi }^{2}=1  \label{c7.3}
\end{equation}
and angular brackets mean the averaging over the argument $\omega t$.
Substituting relations (\ref{c7.2}) and (\ref{c7.3}) in the relation (\ref%
{c7.1}) and taking into account that 
\begin{equation}
\left\langle \cos ^{2}\left( \omega t\right) \right\rangle =\left\langle
\sin ^{2}\left( \omega t\right) \right\rangle =\frac{1}{2},\qquad
\left\langle \sin \left( \omega t\right) \cos \left( \omega t\right)
\right\rangle =0  \label{c7.4}
\end{equation}
we obtain 
\begin{equation}
\left\langle -\frac{e}{c}\mathbf{A\dot{x}}\right\rangle +\frac{e}{c}\mathbf{A%
}\frac{\mathbf{p}}{\kappa _{0}m}=+\frac{e}{2c\omega }V_{\alpha }A_{\alpha
,\mu }\varepsilon _{\mu \nu \sigma }\xi _{\nu }V_{\sigma }-\frac{e}{2c\omega 
}A_{\alpha ,\mu }\varepsilon _{\alpha \beta \gamma }V_{\mu }\xi _{\beta
}V_{\gamma }  \label{c7.5}
\end{equation}

In vector form the expression (\ref{c7.5}) takes the form 
\begin{equation}
\left\langle -\frac{e}{c}\mathbf{A\dot{x}}\right\rangle +\frac{e}{c}\mathbf{A%
}\frac{\mathbf{p}}{\kappa _{0}m}=-\frac{\kappa _{0}e\hbar }{4mc^{3}}\mathbf{V%
}\left( \left( \left( \mathbf{\xi \times V}\right) \cdot \mathbf{\nabla }%
\right) \mathbf{A}-\nabla \left( \mathbf{A\cdot }\left( \mathbf{\xi \times V}%
\right) \right) \right)  \label{c7.6}
\end{equation}

Let us transform the last term in rhs of (\ref{c7.6}) by means of the vector
formula 
\begin{equation*}
\mathbf{\nabla }\left( \mathbf{F\cdot G}\right) =\left( \mathbf{F\cdot
\nabla }\right) \mathbf{G}+\left( \mathbf{G\cdot \nabla }\right) \mathbf{F}+%
\mathbf{F\times }\left( \mathbf{\nabla \times G}\right) +\mathbf{G\times }%
\left( \mathbf{\nabla \times F}\right)
\end{equation*}

Setting $\mathbf{F}=\mathbf{A}$, $\mathbf{G}=\mathbf{\xi \times V}$ and
taking into account, that $\mathbf{\xi }=$const, $\mathbf{V}=$const, $\left( 
\mathbf{\xi V}\right) =0$, we obtain instead of (\ref{c7.6}) 
\begin{eqnarray}
\left\langle -\frac{e}{c}\mathbf{A\dot{x}}\right\rangle &=&-\frac{e}{c}%
\mathbf{A}\frac{\mathbf{p}}{\kappa _{0}m}+\frac{\kappa _{0}e\hbar }{4mc^{3}}%
\mathbf{V}\left( \left( \mathbf{\xi \times V}\right) \times \left( \mathbf{%
\nabla \times A}\right) \right)  \notag \\
&=&-\frac{e}{c}\mathbf{A}\frac{\mathbf{p}}{\kappa _{0}m}+\frac{\kappa
_{0}e\hbar }{4mc^{3}}\mathbf{V}\left( \left( \mathbf{\xi \times V}\right)
\times \mathbf{H}\right)  \notag \\
&=&-\frac{e}{c}\mathbf{A}\frac{\mathbf{p}}{\kappa _{0}m}-\frac{\kappa
_{0}e\hbar }{4mc}\left( \frac{\mathbf{V}}{c}\right) ^{2}\left( \mathbf{\xi H}%
\right)  \label{c7.7}
\end{eqnarray}
where 
\begin{equation*}
\mathbf{H}=\mathbf{\nabla \times A}
\end{equation*}
is the magnetic field.

After averaging the high frequency relativistic term turns into 
\begin{equation}
\left\langle \frac{\hbar }{4c^{2}}(\dot{\mathbf{x}}\times \ddot{\mathbf{x}})%
\mathbf{\xi }\right\rangle \rightarrow \kappa _{0}\frac{m\mathbf{V}^{2}}{2}=%
\text{const}  \label{c7.7a}
\end{equation}%
This term is constant, and it does not contribute to the dynamic equations.

Let us substitute (\ref{c7.7}) in the action (\ref{b3.11}). Taking into
account, that the average low frequency velocity $\mathbf{\dot{x}}=\frac{%
\mathbf{p}}{\kappa _{0}m}$\textbf{, }we obtain instead of (\ref{b3.11}) 
\begin{equation}
\mathcal{A}_{\mathrm{Dcl}}[\mathbf{x},\mathbf{\xi }]=\int \left\{ \frac{1}{2}%
\kappa _{0}m\mathbf{\dot{x}}^{2}-\frac{e}{c}A_{0}-\frac{e}{c}\mathbf{A\dot{x}%
}-\frac{\kappa _{0}e\hbar }{4mc}\left( \frac{\mathbf{V}}{c}\right)
^{2}\left( \mathbf{\xi H}\right) +\hbar {\frac{(\dot{\mathbf{\xi }}\times 
\mathbf{\xi })\mathbf{z}}{2(1+\mathbf{\xi z})}}\right\} dt  \label{c7.8}
\end{equation}

The action (\ref{c7.8}) differs from the action for the classical Pauli
particle in the sense that it depends on the free parameter $V$. This
parameter describes the intensity of the high frequency rotation. To obtain
the action for the classical Pauli particle we should identify the variable $%
\mathbf{\xi }$ with the particle spin and set $V=c\sqrt{2}$. Although this
value of the velocity is relativistic and unreal, the contribution of the
high frequency term in the action (\ref{c7.8}) has the same form as in the
action for the classical Pauli particle.

Note that taking for averaging the relativistical expressions (\ref{f2.28}),
or (\ref{f2.31}) for $V$ and $\omega $, we obtain instead of $V=c\sqrt{2}$
another expressions, where $\left\vert V\right\vert <c$.

\section{Concluding remarks}

This paper is devoted to comparison of the Newtonian strategy and the
experimental-fitting strategy in their application to the microcosm
investigation. In this comparison we underline the role of mistakes in the
foundation of physical theory. These mistakes are actual only in the theory
of microcosm phenomena. It was the mistakes, that have lead to a replacement
of the Newtonian strategy, dominating in 19th century, by the Ptolemaic
experimental-fitting approach, dominating in microcosm investigations of
20th century. The Ptolemaic approach and experimental-fitting way of
thinking are the main obstacles on the path of development of the
satisfactory fundamental microcosm theory. The Ptolemaic approach works very
well at investigation of concrete physical phenomena, because it is
insensitive to mistakes in the foundation of the theory. However, it is not
adequate for construction of a fundamental physical theory, because it
create only list of prescriptions, but not a logical structure. Extension of
a fundamental theory to the new region of relativistic microcosm phenomena
is produced easier, if the theory is a logical structure, but not a list of
prescriptions. Discovery and correction of mistakes is the only way for
construction of a logical structure instead of the Ptolemaic list of
prescriptions. In such a situation it is very important to distinguish
between a mistake and a simple deficiency of our knowledge, as well as
between the mistake and incorrect hypothesis. A mistake is an incorrect
information, whereas a deficiency of knowledge is simply a lack of
information. A hypothesis may be correct in some situation and invalid in
other situation. A hypothesis may be justified or removed, because it lies
outside the logical structure of the satisfactory theory. On the contrary,
the mistake must be discovered and corrected. It is contained in the logical
structure of the satisfactory theory and it may not be ignored.
Unfortunately, the experimental-fitting approach does not distinguish
between a mistake and an incorrect hypothesis, because at this approach a
theory is a list of prescriptions having equal importance, but not a logical
structure.

The most contemporary researchers of microcosm are educated in the Ptolemaic
experimental-fitting approach. The Newtonian approach is unknown for them,
although it is not a new approach. They believe that any theory, which
explain experimental data, is a good theory. Of course, the
experimental-fitting approach has an historical reason, but, when the main
mistakes are discovered and corrected, the list of prescriptions can and
must be replaced by a logical structure. Now there is no reason for a use of
the Ptolemaic approach for construction of the theory of microcosm phenomena.

Rejecting the quantum principles and using dynamical methods of
investigation, we obtain results, which cannot be obtained by means of the
conventional technique, based on the axiomatic representation of the quantum
mechanics. In particular, dynamical methods lead to such results: (1)
formalization of the procedure of transition to classical approximation, (2)
composite structure of the Dirac particle and (3) nonrelativistic
description of the internal degrees of freedom of the Dirac particle. These
results cannot be obtained by conventional methods of investigation.

Mistakes in the foundation of a theory are rather specific. This is not
logical, or mathematical mistakes. These mistakes are associative delusions,
where one uses incorrect associations between our ideas on the properties of
the phenomena of the real world. It is rather difficult to discover the
associative delusions. We illustrate this in the example of discovery of the
nonrelativistic character of the Dirac equation.

The Dirac equation can be written in the relativistically covariant form. It
is common practice to think, that it means that the Dirac equation describes
relativistical processes and has the Lorentz symmetry, i.e. the set of all
its solutions is transformed to the same set of solutions at any Lorentz
transformation. This opinion has been existing for many years, and we try to
understand the reason of this viewpoint.

The relativistic character of dynamic equations associates with the
representation of these equations in the relativistically covariant form.
However, this association is valid only at some additional conditions, which
are fulfilled practically always, and as a result these conditions are not
mentioned usually in the conventional formulation of the relativistic
invariance (compatibility of dynamic equations with the principles of
relativity). Unfortunately, in the case of the Dirac equation these
additional conditions are not fulfilled, and the Dirac equation appears to
be formally nonrelativistic. In reality, only internal degrees of freedom
are nonrelativistic. If these internal degrees of freedom are ignored, the
Dirac particle appears to be relativistic.

The additional constraint in the formulation of the relativistical
invariance changes the formulation. The correct formulation looks as
follows. \textit{Symmetry of dynamic equations, written in the
relativistically covariant form coincides with the symmetry of their
absolute objects} \cite{A67}. The absolute objects are such quantities,
which are the same for all solutions. Usually such an absolute object is the
metric tensor, which has the form $g_{ik}=$diag$\left\{
c^{2},-1,-1,-1\right\} $. The group of symmetry of $g_{ik}$ is the Lorentz
group, and the symmetry group of dynamic equations appears to be the Lorentz
group. The Maxwell equations, the Klein-Gordon equation and many other
dynamic equations for real dynamic systems contain only the metric tensor as
an absolute object, and the formulation of relativistical invariance is
simplified. It looks as follows. \textit{The symmetry group of dynamic
equations, written in the relativistically covariant form is the Lorentz
group}. In such a form it used by most researchers.

The Dirac equation does not contain the metric tensor. Instead it contains
the $\gamma $-matrices $\gamma ^{i}$, $i=0,1,2,3$. The $\gamma $-matrices
form a matrix 4-vector, whose symmetry group is lower, than the Lorentz
group. As a result the Dirac equation appears to have not a symmetry of the
Lorentz group. In other words, the Dirac equation appears to be
nonrelativistic equation.

What physical situation is behind this result? Why does the dynamic
equation, written in the relativistically covariant form, become to be
nonrelativistic, if it contains an absolute vector? To answer this question,
we consider an example of a charged classical particle, moving in the given
electromagnetic field $F^{ik}$.

Dynamic equation for the relativistic particle may written in the
noncovariant form%
\begin{equation}
\frac{d}{dt}\frac{m\dot{x}^{\mu }}{\sqrt{1-\frac{\mathbf{\dot{x}}^{2}}{c^{2}}%
}}=\frac{e}{c}F^{\mu 0}+\frac{e}{c}F^{\mu \nu }g_{\nu \beta }\dot{x}^{\beta
},\qquad \mu =1,2,3,\qquad \mathbf{\dot{x}}\equiv \frac{d\mathbf{x}}{dt}
\label{a9.1}
\end{equation}%
and in the relativistically covariant form%
\begin{equation}
m\frac{d^{2}x^{k}}{d\tau ^{2}}=\frac{e}{c}F^{kl}g_{ls}\frac{dx^{s}}{d\tau }%
,\qquad k=0,1,2,3  \label{a9.2}
\end{equation}%
where $\tau $ is the proper time, $e$, $m$ are respectively the particle
charge and the particle mass.

If the particle is nonrelativistic the dynamic equation is written in the
noncovariant form%
\begin{equation}
m\frac{d^{2}x^{\mu }}{dt^{2}}=\frac{e}{c}F^{\mu 0}+\frac{e}{c}F^{\mu \nu
}g_{\nu \beta }\frac{dx^{\beta }}{dt},\qquad \mu =1,2,3  \label{a9.3}
\end{equation}%
Can the dynamic equations (\ref{a9.3}) for the nonrelativistic particle be
written in the relativistically covariant form? The answer is yes, although
most researchers believe that it is impossible. In the relativistically
covariant form the dynamic equations (\ref{a9.3}) have the form 
\begin{equation}
m\frac{d}{d\tau }\left[ \left( l_{k}\dot{x}^{k}\right) ^{-1}\dot{x}^{i}-{%
\frac{1}{2}}g^{ik}l_{k}\left( l_{j}\dot{x}^{j}\right) ^{-2}\dot{x}^{s}g_{sl}%
\dot{x}^{l}\right] =\frac{e}{c}F^{il}g_{lk}\dot{x}^{k};\qquad i=0,1,2,3
\label{a9.4}
\end{equation}%
where $\dot{x}^{k}\equiv dx^{k}/d\tau $. The quantity $l_{k}$, $k=0,1,2,3$
is a constant timelike unit 4-vector 
\begin{equation}
g^{ik}l_{i}l_{k}=1;  \label{a9.5}
\end{equation}%
Using the special choice of $l_{k}=\left\{ c,0,0,0\right\} $ and
substituting it in (\ref{a9.4}), it is easy to verify, that we obtain the
dynamic equations (\ref{a9.3}) for $i=1,2,3$. For $i=0$ we obtain dynamic
equation, which is a corollary of (\ref{a9.3}).

As far as dynamic equations for both relativistic and nonrelativistic
particles can be written in the noncovariant form and in the
relativistically covariant one, it is clear that the difference between the
relativistic and nonrelativistic descriptions is not connected with form of
dynamic equations. There is anything else, which distinguishes the
relativistic conception from the nonrelativistic one. It is well known that
the difference lies in different space-time conceptions. In the Newtonian
conception there is an absolute simultaneity and there are two invariant
quantities: absolute time $t$ and absolute space distance $r$, whereas in
the relativistic space-time conception there exists only one absolute
quantity: the space-time interval $s=\sqrt{c^{2}t^{2}-r^{2}}$. The Newtonian
space-time $\mathcal{S}_{\mathrm{N}}$ has seven-parametric continuous group
of motion, whereas the Minkowski space-time $\mathcal{S}_{\mathrm{M}}$ has
ten-parametric continuous group of motion. Besides, the Newtonian space-time 
$\mathcal{S}_{\mathrm{N}}$ may be considered to be the Minkowski space-time $%
\mathcal{S}_{\mathrm{M}}$ with additional geometric structure $\mathcal{L}$,
given in it. In other words, $\mathcal{S}_{\mathrm{N}}=\mathcal{S}_{\mathrm{M%
}}\wedge \mathcal{L}$. The additional structure $\mathcal{L}$ is a specific
timelike direction in $\mathcal{S}_{\mathrm{M}}$, described by the constant
timelike vector $l_{k}$. Introduction of $\mathcal{L}$ admits one to
construct two invariants in $\mathcal{S}_{\mathrm{M}}\wedge \mathcal{L}$ 
\begin{equation}
t=l_{k}x^{k},\qquad r=\sqrt{x^{k}x_{k}+\left( l_{k}x^{k}\right) ^{2}}
\label{a9.6}
\end{equation}%
for a vector $x^{k}$, whereas in $\mathcal{S}_{\mathrm{M}}$ we have only one
invariant $s=\sqrt{x^{k}x_{k}}$. It does not refer to $\mathcal{L}$.

The Newtonian space-time $\mathcal{S}_{\mathrm{N}}$ considered as $\mathcal{S%
}_{\mathrm{M}}\wedge \mathcal{L}$ admits only such motions of $\mathcal{S}_{%
\mathrm{M}}$, which transform vector $l_{k}$ into the same vector $l_{k}$
and do not violate the structure $\mathcal{L}$. The condition of the
structure $\mathcal{L}$ conservation at the space-time motion reduces the
ten-parametric group of motion of $\mathcal{S}_{\mathrm{M}}$ to
seven-parametric group of motion of $\mathcal{S}_{\mathrm{M}}\wedge \mathcal{%
L}$. In general, at the relativistically covariant description the absolute
objects, introduced by Anderson \cite{A67} may be considered as the
quantities, describing additional structures in $\mathcal{S}_{\mathrm{M}}$.
It means, that any system of dynamic equations may be written in the
relativistically covariant form, provided the proper absolute objects
(additional structures) are introduced. Thus, to determine, whether the
dynamic equations are compatible with the principles of relativity, we may
write them in the relativistically covariant form and determine whether or
not they contain absolute objects and what are properties of these absolute
objects. If the dynamic equations contain the constant timelike vector $%
l_{k} $, we have nonrelativistic dynamic system, because $l_{k}$ describes
the additional space-time structure, characteristic for the Newtonian space $%
\mathcal{S}_{\mathrm{N}}$ represented as $\mathcal{S}_{\mathrm{M}}\wedge 
\mathcal{L}$.

Such an approach is convenient in the sense, that it does not contain a
reference to the coordinate system, which is simply a method of description.
Relativistic character of dynamic equation is connected directly with
absence of additional space-time structures in $\mathcal{S}_{\mathrm{M}}$,
but not with the relativistically covariant form of the dynamic equations,
because any dynamic equations can be always written in the relativistically
covariant form, provided the proper geometrical structure is introduced in $%
\mathcal{S}_{\mathrm{M}}$ . The relativistically covariant dynamic equation
is relativistic, provided it does not contain a reference to some additional
structure. However, such a formulation is unreliable, because the
reservation of a reference to additional structure may be omitted by
mistake. In this case the relativistic character of dynamic equations appear
to be connected with the relativistic covariance of these equations, but not
with the additional structure $\mathcal{L}$ in $\mathcal{S}_{\mathrm{M}}$.
It is this case that takes place in reality. As a result we have an
associative mistake, when the relativistic invariance is associated with the
relativistic covariance, although in reality the relativistic invariance is
associated with an absence of additional geometrical structures in $\mathcal{%
S}_{\mathrm{M}}$.

The experimental-fitting style of investigation, applied everywhere, is the
main defect of the contemporary investigation strategy. This style is
applied not only in investigation of concrete physical phenomena of
microcosm, where its application is admissible. It is applied also at
construction of the fundamental physical theory, that is not admissible,
because the fundamental physical theory is a systematization of our
knowledge and establishment of logical connection between the fundamental
concepts. The fundamental physical theory is a logical structure, but not a
list of rules, which should be used for explanation experimental data. The
list of the rules may contain the rules, which are contradictory between
themselves, but the statements of a logical structure must not be
contradictory.

Unfortunately, the Newtonian investigation strategy is not used practically,
because it cannot be used, if our fundamental concepts contain mistakes.
Some of them were listed in introduction. These mistakes were not actual in
the 19th century, when the microcosm was not investigated, and the Newtonian
investigation strategy was dominating. Existence of these mistakes during
the 20th century was the reason, why at the microcosm investigations the
Newtonian strategy was replaced by the experimental-fitting investigation
strategy. The last has the advantage, that it is insensitive to mistakes in
the foundation of a physical theory. Following the experimental-fitting
strategy, the researcher of microcosm did not try to find mistakes of their
predecessors. Further more, doubts in results, obtained by great
predecessors were considered to be a bad form, generated by self-conceit.
Instead of searching for mistakes, that was prescribed by the Newtonian
strategy, researchers invented new hypotheses. \textit{Three generations} of
the microcosm researchers were educated in ideas of the experimental-fitting
strategy. Having investigated the Dirac equation and obtaining the first
result, that the Pauli particle is the nonrelativistic approximation of the
Dirac particle, they did not try to carry out the investigation completely,
because the experimental-fitting strategy does not demand this. Why is it
necessary, if the obtained result explains the experimental data? The fact
that the investigation is not complete, and the Dirac particle is a
composite particle was not considered, although such incomplete
investigation was incorrect mathematically, and the mistake could be found
at the scrupulous mathematical investigation.

Preconceptions of the experimental-fitting style of thinking are very
strong. Even the author of this paper is not free of them, although he is an
adherent of the Newtonian strategy and tries to use this strategy in his
investigations. For instance, he could not find the mistake in the
nonrelativistic approximation of the Dirac equation. He has paid attention
on the reduction of the order of the dynamic system and on the small
parameter before the highest derivatives only after he had discovered, that
the Dirac particle was composite. The last result was obtained from other
consideration \cite{R2004,R2004a}.

\bigskip

\renewcommand{\theequation}{\Alph{section}.\arabic{equation}} %
\renewcommand{\thesection}{\Alph{section}} \setcounter{section}{0} %
\centerline{\Large \bf Appendices}

\section{Transformation of the action for the statistical ensemble}

Let us transform the action 
\begin{equation}
\mathcal{E}\left[ \mathcal{S}_{\mathrm{st}}\right] :\qquad \mathcal{A}_{%
\mathcal{E}\left[ \mathcal{S}_{\mathrm{st}}\right] }\left[ \mathbf{x},%
\mathbf{u}_{\mathrm{st}}\right] \mathbf{=}\int \left\{ \frac{m\mathbf{\dot{x}%
}^{2}}{2}-\frac{e}{c}A_{0}-\frac{e}{c}\mathbf{A}\frac{d\mathbf{x}}{dt}+\frac{%
m\mathbf{u}_{\mathrm{st}}^{2}}{2}-\frac{\hbar }{2}\mathbf{\nabla u}_{\mathrm{%
st}}\right\} dtd\mathbf{\xi }  \label{g1.19}
\end{equation}
for the statistical ensemble of stochastic particles, moving in the given
electromagnetic field $A=\left\{ A_{0},\mathbf{A}\right\} =\left\{
A_{0},A_{1},A_{2},A_{3}\right\} $. Here $\mathbf{x}=\mathbf{x}\left( t,%
\mathbf{\xi }\right) $, $\mathbf{u}_{\mathrm{st}}=\mathbf{u}_{\mathrm{st}%
}\left( t,\mathbf{x}\right) $ are dependent dynamic variables, and $\mathbf{%
\nabla =}\left\{ \partial _{1},\partial _{2},\partial _{3}\right\} \mathbf{=}%
\left\{ \frac{\partial }{\partial x^{1}},\frac{\partial }{\partial x^{2}},%
\frac{\partial }{\partial x^{3}}\right\} $. The variable $\mathbf{x}$
describes the regular component of the stochastic particle motion. The
dynamic variable $\mathbf{u}_{\mathrm{st}}$ is a function of $t,\mathbf{x}$
and depends on $\mathbf{\xi }$ via $\mathbf{x}$. The quantity $\mathbf{u}_{%
\mathrm{st}}$ may be regarded as the mean velocity of the stochastic
component, whereas $\mathbf{x}=\mathbf{x}\left( t,\mathbf{\xi }\right) $
describes the regular component of the particle motion. The last term in (%
\ref{g1.19}) describes influence of the stochasticity on the regular
evolution component.

To eliminate variable $\mathbf{u}_{\mathrm{st}}$, we should to solve dynamic
equations $\delta \mathcal{A}/\mathcal{\delta }\mathbf{u}_{\mathrm{st}}=0$
with respect to $\mathbf{u}_{\mathrm{st}}$. As far as $\mathbf{u}_{\mathrm{st%
}}$ is a function of $t,\mathbf{x}$, we should go to independent variables $%
t,\mathbf{x}$ in the action (\ref{g1.19}). We obtain instead of (\ref{g1.19}%
) 
\begin{equation}
\mathcal{A}_{\mathcal{E}\left[ \mathcal{S}_{\mathrm{st}}\right] }\left[ 
\mathbf{\xi },\mathbf{u}_{\mathrm{st}}\right] \mathbf{=}\int \left\{ \frac{m%
\mathbf{\dot{x}}^{2}}{2}-\frac{e}{c}A_{0}-\frac{e}{c}\mathbf{A}\frac{d%
\mathbf{x}}{dt}+\frac{m\mathbf{u}_{\mathrm{st}}^{2}}{2}-\frac{\hbar }{2}%
\mathbf{\nabla u}_{\mathrm{st}}\right\} \rho \left( t,\mathbf{x}\right) dtd%
\mathbf{x}  \label{g1.20}
\end{equation}
where $\mathbf{\ \xi }$, $\mathbf{u}_{\mathrm{st}}$ are dependent variables,
whereas $t,\mathbf{x}$ are independent variables. Here $\rho $ and $\mathbf{%
\dot{x}}=\mathbf{u}$ are functions of $\mathbf{\xi }$, defined by the
relations 
\begin{equation}
\rho =\frac{\partial \left( \xi _{1},\xi _{2},\xi _{3}\right) }{\partial
\left( x^{1},x^{2},x^{3}\right) },\qquad \mathbf{\dot{x}}\equiv \mathbf{u}%
\equiv \frac{\partial \left( \mathbf{x,}\xi _{1},\xi _{2},\xi _{3}\right) }{%
\partial \left( t,\xi _{1},\xi _{2},\xi _{3}\right) }=\frac{1}{\rho }\frac{%
\partial \left( \mathbf{x,}\xi _{1},\xi _{2},\xi _{3}\right) }{\partial
\left( t,x^{1},x^{2},x^{3}\right) },  \label{g1.21}
\end{equation}
Variation of (\ref{g1.20}) with respect $\mathbf{u}_{\mathrm{st}}$ gives 
\begin{equation}
\frac{\delta \mathcal{A}_{\mathcal{E}\left[ \mathcal{S}_{\mathrm{st}}\right]
}}{\mathcal{\delta }\mathbf{u}_{\mathrm{st}}}=m\mathbf{u}_{\mathrm{st}}\rho +%
\frac{\hbar }{2}\mathbf{\nabla }\rho =0  \label{g1.22}
\end{equation}
Resolving the equation (\ref{g1.22}) with respect to $\mathbf{u}_{\mathrm{st}%
}$ in the form 
\begin{equation}
\mathbf{u}_{\mathrm{st}}=-\frac{\hbar }{2m}\mathbf{\nabla }\ln \rho ,
\label{g1.23}
\end{equation}
we obtain instead of (\ref{g1.20}) 
\begin{equation}
\mathcal{A}_{\mathcal{E}\left[ \mathcal{S}_{\mathrm{st}}\right] }\left[ 
\mathbf{\xi }\right] \mathbf{=}\int \left\{ \frac{m}{2}\left( \frac{d\mathbf{%
x}}{dt}\right) ^{2}-\frac{e}{c}A_{0}-\frac{e}{c}\mathbf{A}\frac{d\mathbf{x}}{%
dt}-\frac{\hbar ^{2}}{8m}\frac{\left( \mathbf{\nabla }\rho \right) ^{2}}{%
\rho ^{2}}\right\} \rho dtd\mathbf{x}  \label{g1.24}
\end{equation}
where $\rho $ and $\frac{d\mathbf{x}}{dt}$ are functions of space-time
derivatives of $\mathbf{\xi }=\left\{ \xi _{1},\xi _{2},\xi _{3}\right\} $,
determined by the relations (\ref{g1.21}). The action (\ref{g1.24})
describes some ideal charged fluid with the internal energy per unit mass 
\begin{equation}
U\left( \rho ,\mathbf{\nabla }\rho \right) =\frac{\hbar ^{2}}{8m}\frac{%
\left( \mathbf{\nabla }\rho \right) ^{2}}{\rho ^{2}}  \label{g1.25}
\end{equation}

Let us introduce new dependent variables $j=\left\{ \rho ,\rho \mathbf{u}%
\right\} =\left\{ j^{k}\right\} ,\;\;k=0,1,2,3$ by means of relations (\ref%
{g1.21}). From formal viewpoint it is convenient to represent the
hydrodynamic variables $j=\left\{ \rho ,\rho \mathbf{u}\right\} =\left\{
j^{k}\right\} $, $k=0,1,2,3$ in the form 
\begin{equation}
j^{k}=\frac{\partial \left( x^{k}\mathbf{,}\xi _{1},\xi _{2},\xi _{3}\right) 
}{\partial \left( x^{0},x^{1},x^{2},x^{3}\right) }=\frac{\partial J}{%
\partial \xi _{0,k}},\qquad k=0,1,2,3  \label{g1.26}
\end{equation}
where the Jacobian 
\begin{equation}
J=\frac{\partial \left( \xi _{0}\mathbf{,}\xi _{1},\xi _{2},\xi _{3}\right) 
}{\partial \left( x^{0},x^{1},x^{2},x^{3}\right) }=\det \left\vert
\left\vert \xi _{i,k}\right\vert \right\vert ,\qquad \xi _{l,k}\equiv
\partial _{k}\xi _{l},\qquad l,k=0,1,2,3  \label{g1.27}
\end{equation}
is considered to be a function of variables $\xi _{l,k}\equiv \partial
_{k}\xi _{l},\;l,k=0,1,2,3$. The variable $\xi _{0}$ is the new dependent
variable (temporal Lagrangian coordinate), which appears to be fictitious.

We introduce new dynamic variables by the Lagrange multipliers $p=\left\{
p_{k}\right\} ,\;\;k=0,1,2,3$, and obtain instead of (\ref{g1.24}) 
\begin{equation}
\mathcal{A}_{\mathcal{E}\left[ \mathcal{S}_{\mathrm{st}}\right] }\left[ \xi 
\mathbf{,}j,p\right] \mathbf{=}\int \left\{ \frac{m}{2\rho }j^{\alpha
}j^{\alpha }-\frac{e}{c}A_{0}\rho -\frac{e}{c}A_{\alpha }j^{\alpha
}-p_{k}\left( j^{k}-\frac{\partial J}{\partial \xi _{0,k}}\right) -\frac{%
\hbar ^{2}}{8m}\frac{\left( \mathbf{\nabla }\rho \right) ^{2}}{\rho }%
\right\} d^{4}x  \label{g1.28}
\end{equation}%
where $\xi =\left\{ \xi _{k}\right\} $,$\;\;k=0,1,2,3$.

Variation of the action (\ref{g1.28}) with respect to $\xi _{l}$ leads to
the dynamic equations 
\begin{equation}
\frac{\delta \mathcal{A}_{\mathcal{E}\left[ \mathcal{S}_{\mathrm{st}}\right]
}}{\mathcal{\delta \xi }_{l}}=\partial _{s}\left( p_{k}\frac{\partial ^{2}J}{%
\partial \xi _{0,k}\partial \xi _{l,s}}\right) =0,\qquad l=0,1,2,3
\label{g1.29}
\end{equation}
As far as the variable $\xi _{0}$ is fictitious, there are only three
independent equations among four equations (\ref{g1.29}).

Using identities 
\begin{equation}
\frac{\partial ^{2}J}{\partial \xi _{0,k}\partial \xi _{l,s}}\equiv
J^{-1}\left( \frac{\partial J}{\partial \xi _{0,k}}\frac{\partial J}{%
\partial \xi _{l,s}}-\frac{\partial J}{\partial \xi _{0,s}}\frac{\partial J}{%
\partial \xi _{l,k}}\right)  \label{A.9}
\end{equation}
\begin{equation}
\frac{\partial J}{\partial \xi _{i,l}}\xi _{k,l}\equiv J\delta
_{k}^{i},\qquad \partial _{l}\frac{\partial ^{2}J}{\partial \xi
_{0,k}\partial \xi _{i,l}}\equiv 0  \label{A.10}
\end{equation}
and designations (\ref{g1.26}), we can eliminate the variables $\mathbf{\xi }
$ from the equations (\ref{g1.29}). We obtain 
\begin{equation}
j^{k}\partial _{l}p_{k}-j^{k}\partial _{k}p_{l}=0,\qquad l=0,1,2,3
\label{A.10a}
\end{equation}

Variation of (\ref{g1.28}) with respect to $j^{\beta }$ and $j^{0}=\rho $
gives respectively 
\begin{equation}
p_{\beta }=m\frac{j^{\beta }}{\rho }-\frac{e}{c}A_{\beta },\qquad \beta
=1,2,3  \label{A.17}
\end{equation}
\begin{equation}
p_{0}=-\frac{m}{2\rho ^{2}}j^{\alpha }j^{\alpha }-\frac{e}{c}A_{0}+\frac{%
\hbar ^{2}}{8m}\left( 2\frac{\mathbf{\nabla }^{2}\rho }{\rho }-\frac{\left( 
\mathbf{\nabla }\rho \right) ^{2}}{\rho ^{2}}\right)  \label{A.10b}
\end{equation}
Eliminating $p_{k}$ from the equations (\ref{A.10a}) by means of relations (%
\ref{A.17}), (\ref{A.10b}), we obtain hydrodynamic equations for the ideal
charged fluid in the conventional form 
\begin{equation}
\left( \partial _{0}+v^{\alpha }\partial _{\alpha }\right) v^{\mu }=\frac{e}{%
mc}F_{\mu 0}+\frac{e}{mc}F_{\mu \alpha }v^{\alpha }-\frac{1}{m\rho }\partial
_{\mu }p,\qquad \mu =1,2,3  \label{A.10c}
\end{equation}
where the pressure $p$ and the electromagnetic field $F_{ik}$ are defined by
the relations 
\begin{equation}
p=\frac{\hbar ^{2}}{8m}\left( \frac{\left( \mathbf{\nabla }\rho \right) ^{2}%
}{\rho ^{2}}-2\frac{\mathbf{\nabla }^{2}\rho }{\rho }\right) ,\qquad
F_{ik}=\partial _{k}A_{i}-\partial _{i}A_{k},\qquad i,k=0,1,2,3
\label{A.10d}
\end{equation}

The wave function is constructed of potentials. The equations (\ref{A.10c})
does not contain potentials $\mathbf{\xi }$ and $A_{k}$, and they cannot be
used for description of the fluid in terms of the wave function. To
construct a description in terms of the wave function, we should not to
eliminate potentials $\mathbf{\xi }$ from the equations (\ref{g1.29}).
Instead, we should integrate them. The dynamic equations (\ref{g1.29}) may
be considered to be linear partial differential equations with respect to
variables $p_{k}$. They can be solved in the form 
\begin{equation}
p_{k}=b\left( \partial _{k}\varphi +g^{\alpha }\left( \mathbf{\xi }\right)
\partial _{k}\xi _{\alpha }\right) ,\qquad k=0,1,2,3  \label{g1.30}
\end{equation}%
where $g^{\alpha }\left( \mathbf{\xi }\right) ,\;\;\alpha =1,2,3$ are
arbitrary functions of the argument $\mathbf{\xi }=\left\{ \xi _{1},\xi
_{2},\xi _{3}\right\} $, $b$ is an arbitrary real constant, and $\varphi $
is the variable $\xi _{0}$, which ceases to be fictitious.

One can test by the direct substitution that the relation (\ref{g1.30}) is
the general solution of linear equations (\ref{g1.29}). Indeed, using (\ref%
{A.9}) and the second identity (\ref{A.10}), the equations (\ref{g1.29}) may
be written in the form 
\begin{equation}
\frac{\partial ^{2}J}{\partial \xi _{0,k}\partial \xi _{l,s}}\partial
_{s}p_{k}=J^{-1}\left( \frac{\partial J}{\partial \xi _{0,k}}\frac{\partial J%
}{\partial \xi _{l,s}}-\frac{\partial J}{\partial \xi _{0,s}}\frac{\partial J%
}{\partial \xi _{l,k}}\right) \partial _{s}p_{k}=0  \label{g1.31}
\end{equation}
Substituting (\ref{g1.30}) in (\ref{g1.31}) and taking into account
antisymmetry of the bracket in (\ref{g1.31}) with respect to indices $k$ and 
$s$, we obtain 
\begin{equation}
J^{-1}\left( \frac{\partial J}{\partial \xi _{0,k}}\frac{\partial J}{%
\partial \xi _{l,s}}-\frac{\partial J}{\partial \xi _{0,s}}\frac{\partial J}{%
\partial \xi _{l,k}}\right) \frac{\partial g^{\alpha }\left( \mathbf{\xi }%
\right) }{\partial \xi _{\mu }}\xi _{\mu ,s}\xi _{\alpha ,k}=0  \label{g1.32}
\end{equation}
The relation (\ref{g1.32}) is the valid equality, as it follows from the
first identity (\ref{A.10}).

Let us substitute (\ref{g1.30}) in the action (\ref{g1.28}). Taking into
account the first identity (\ref{A.10}) and omitting the term 
\begin{equation*}
\frac{\partial J}{\partial \xi _{0,k}}\partial _{k}\varphi =\frac{\partial
\left( \varphi ,\xi _{1},\xi _{2},\xi _{3}\right) }{\partial \left(
x^{0},x^{1},x^{2},x^{3}\right) }
\end{equation*}
which does not contribute to the dynamic equation, we obtain 
\begin{equation}
\mathcal{E}\left[ \mathcal{S}_{\mathrm{st}}\right] :\qquad \mathcal{A}_{%
\mathcal{E}\left[ \mathcal{S}_{\mathrm{st}}\right] }\left[ \varphi ,\mathbf{%
\xi },j\right] =\int \left\{ \frac{m}{2}\frac{j^{\alpha }j^{\alpha }}{j^{0}}-%
\frac{e}{c}A_{k}j^{k}-j^{k}p_{k}-\frac{\hbar ^{2}}{8m}\frac{\left( \mathbf{%
\nabla }\rho \right) ^{2}}{\rho }\right\} d^{4}x\mathbf{,}  \label{A.12}
\end{equation}
Here quantities $p_{k}$ are determined by the relations (\ref{g1.30}).

The action in the form (\ref{A.12}) is remarkable in the sense, that it
contains information on initial values of the fluid velocities $\mathbf{v}=%
\mathbf{j}/\rho $. Dynamic equations, generated by the action (\ref{A.12}),
are partial differential equations, and one needs to give initial values for
variables $\varphi ,\mathbf{\xi }$\textbf{.} But initial values for
variables $\varphi ,\mathbf{\xi }$ determine only labelling of the fluid
particles, and they may be chosen universal. For instance, we may choose for
all fluid flows 
\begin{equation}
\varphi \left( 0,\mathbf{x}\right) =\varphi _{\mathrm{in}}\left( \mathbf{x}%
\right) =0,\qquad \mathbf{\xi }\left( 0,\mathbf{x}\right) =\mathbf{\xi }_{%
\mathrm{in}}\left( \mathbf{x}\right) =\mathbf{x}  \label{A.12b}
\end{equation}
Then the functions $\mathbf{g}\left( \mathbf{\xi }\right) $ are determined
by the initial values of the velocity $\mathbf{v}\left( 0,\mathbf{x}\right) =%
\mathbf{v}_{\mathrm{in}}\left( \mathbf{x}\right) $ in the form \cite{R99} 
\begin{equation}
\mathbf{g}\left( \mathbf{\xi }\right) =\mathbf{v}_{\mathrm{in}}\left( 
\mathbf{\xi }\right)  \label{A.12c}
\end{equation}
The initial value $\rho \left( 0,\mathbf{x}\right) =\rho _{\mathrm{in}%
}\left( \mathbf{x}\right) $ of the density $\rho $ may be also included in
the action (\ref{A.12}). It is necessary only to redefine the connection
between the quantities $j^{k}$ and $\mathbf{\xi }$\textbf{, }substituting
the relations (\ref{g1.26}) by the relations \cite{R99} 
\begin{equation}
j^{k}=\rho _{0}\left( \mathbf{\xi }\right) \frac{\partial \left( x^{k}%
\mathbf{,}\xi _{1},\xi _{2},\xi _{3}\right) }{\partial \left(
x^{0},x^{1},x^{2},x^{3}\right) },\qquad k=0,1,2,3  \label{A.12d}
\end{equation}
where $\rho _{0}\left( \mathbf{\xi }\right) $ is an arbitrary function of $%
\mathbf{\xi }$. At the initial conditions (\ref{A.12c}) this arbitrary
function is to be chosen in the form 
\begin{equation*}
\rho _{0}\left( \mathbf{x}\right) =\rho _{\mathrm{in}}\left( \mathbf{x}%
\right) =\rho \left( 0,\mathbf{x}\right)
\end{equation*}

Now we eliminate the variables $\mathbf{j}=\left\{ j^{1},j^{2},j^{3}\right\} 
$ from the action (\ref{A.12}), using relation (\ref{A.17}). We obtain 
\begin{equation}
\mathcal{A}_{\mathcal{E}\left[ \mathcal{S}_{\mathrm{st}}\right] }\left[ \rho
,\varphi ,\mathbf{\xi }\right] =\int \left\{ -p_{0}-\frac{e}{c}A_{0}-\frac{%
\left( p_{\beta }+\frac{e}{c}A_{\beta }\right) \left( p_{\beta }+\frac{e}{c}%
A_{\beta }\right) }{2m}-\frac{\hbar ^{2}}{8m}\frac{\left( \mathbf{\nabla }%
\rho \right) ^{2}}{\rho ^{2}}\right\} \rho d^{4}x\mathbf{,}  \label{A.18}
\end{equation}%
where the quantities $p_{k}$, $k=0,1,2,3$ are determined by the relation (%
\ref{g1.30}).

Instead of dependent variables $\rho ,\varphi ,\mathbf{\xi }$ we introduce
the $n$-component complex function $\psi =\{\psi _{\alpha }\},\;\;\alpha
=1,2,\ldots ,n$, which is defined by the relations \cite{R99} 
\begin{equation}
\psi _{\alpha }=\sqrt{\rho }e^{i\varphi }u_{\alpha }(\mathbf{\xi }),\qquad
\psi _{\alpha }^{\ast }=\sqrt{\rho }e^{-i\varphi }u_{\alpha }^{\ast }(%
\mathbf{\xi }),\qquad \alpha =1,2,\ldots ,n,  \label{A.19}
\end{equation}
\begin{equation}
\psi ^{\ast }\psi \equiv \sum_{\alpha =1}^{n}\psi _{\alpha }^{\ast }\psi
_{\alpha },  \label{A.20}
\end{equation}
where (*) means the complex conjugate. The quantities $u_{\alpha }(\mathbf{%
\xi })$, $\alpha =1,2,\ldots ,n$ are functions of only variables $\mathbf{%
\xi }$, and satisfy the relations 
\begin{equation}
-\frac{i}{2}\sum_{\alpha =1}^{n}\left( u_{\alpha }^{\ast }\frac{\partial
u_{\alpha }}{\partial \xi _{\beta }}-\frac{\partial u_{\alpha }^{\ast }}{%
\partial \xi _{\beta }}u_{\alpha }\right) =g^{\beta }(\mathbf{\xi }),\qquad
\beta =1,2,3,\qquad \sum_{\alpha =1}^{n}u_{\alpha }^{\ast }u_{\alpha }=1.
\label{A.21}
\end{equation}
The number $n$ is such a natural number that the equations (\ref{A.21})
admit a solution. In general, $n$ depends on the form of the arbitrary
integration functions $\mathbf{g}=\{g^{\beta }(\mathbf{\xi })\}$, $\beta
=1,2,3$. The functions $\mathbf{g}$ determine vorticity of the fluid flow.
If $\mathbf{g}=0$, equations (\ref{A.21}) have the solution $u_{1}=1$, $%
u_{\alpha }=0$, \ $\alpha =2,3,...n$. In this case the function $\psi $ may
have one component, and the fluid flow is irrotational.

In the general case it is easy to verify that 
\begin{equation}
\rho =\psi ^{\ast }\psi ,\qquad \rho p_{0}\left( \varphi ,\mathbf{\xi }%
\right) =-\frac{ib}{2}(\psi ^{\ast }\partial _{0}\psi -\partial _{0}\psi
^{\ast }\cdot \psi )  \label{s5.6}
\end{equation}
\begin{equation}
\rho p_{\alpha }\left( \varphi ,\mathbf{\xi }\right) =-\frac{ib}{2}(\psi
^{\ast }\partial _{\alpha }\psi -\partial _{\alpha }\psi ^{\ast }\cdot \psi
),\qquad \alpha =1,2,3,  \label{s5.7}
\end{equation}
The variational problem with the action (\ref{A.12}) appears to be
equivalent to the variational problem with the action functional 
\begin{eqnarray}
&&\mathcal{A}_{\mathcal{E}\left[ \mathcal{S}_{\mathrm{st}}\right] }[\psi
,\psi ^{\ast }]  \notag \\
&=&\int \left\{ \frac{ib}{2}(\psi ^{\ast }\partial _{0}\psi -\partial
_{0}\psi ^{\ast }\cdot \psi )-\frac{e}{c}A_{0}\rho \right.  \notag \\
&&\left. -\frac{\rho }{2m}\left( -\frac{ib}{2\rho }(\psi ^{\ast }\mathbf{%
\nabla }\psi -\mathbf{\nabla }\psi ^{\ast }\cdot \psi )+\frac{e}{c}\mathbf{A}%
\right) ^{2}-\frac{\hbar ^{2}}{8m}\frac{\left( \mathbf{\nabla }\rho \right)
^{2}}{\rho }\right\} \mathrm{d}^{4}x  \label{s5.8}
\end{eqnarray}
or 
\begin{eqnarray}
&&\mathcal{A}_{\mathcal{E}\left[ \mathcal{S}_{\mathrm{st}}\right] }[\psi
,\psi ^{\ast }]  \notag \\
&=&\int \left\{ \frac{ib}{2}(\psi ^{\ast }\partial _{0}\psi -\partial
_{0}\psi ^{\ast }\cdot \psi )-\frac{e}{c}A_{0}+\frac{b^{2}}{8m\rho }(\psi
^{\ast }\mathbf{\nabla }\psi -\mathbf{\nabla }\psi ^{\ast }\cdot \psi
)^{2}\right.  \notag \\
&&\left. +\frac{ibe}{2mc}\mathbf{A}(\psi ^{\ast }\mathbf{\nabla }\psi -%
\mathbf{\nabla }\psi ^{\ast }\cdot \psi )-\frac{\hbar ^{2}}{8m}\frac{\left( 
\mathbf{\nabla }\rho \right) ^{2}}{\rho }-\frac{\rho }{2m}\left( \frac{e}{c}%
\mathbf{A}\right) ^{2}\right\} \mathrm{d}^{4}x  \label{s5.9}
\end{eqnarray}

For the two-component function $\psi $ \ ($n=2$) the following identity
takes place 
\begin{equation}
(\mathbf{\nabla }\rho )^{2}-(\psi ^{\ast }\mathbf{\nabla }\psi -\mathbf{%
\nabla }\psi ^{\ast }\cdot \psi )^{2}\equiv 4\rho \mathbf{\nabla }\psi
^{\ast }\mathbf{\nabla }\psi -\rho ^{2}\sum\limits_{\alpha =1}^{\alpha
=3}\left( \mathbf{\nabla }s_{\alpha }\right) ^{2},  \label{s5.30}
\end{equation}
\begin{equation}
\rho \equiv \psi ^{\ast }\psi ,\qquad \mathbf{s}\equiv \frac{\psi ^{\ast }%
\mathbf{\sigma }\psi }{\rho },\qquad \mathbf{\sigma }=\{\sigma _{\alpha
}\},\qquad \alpha =1,2,3,  \label{s5.30a}
\end{equation}
where $\sigma _{\alpha }$ are the Pauli matrices. In virtue of the identity (%
\ref{s5.30}) the action (\ref{s5.8}) reduces to the form 
\begin{eqnarray}
&&\mathcal{A}_{\mathcal{E}\left[ \mathcal{S}_{\mathrm{st}}\right] }[\psi
,\psi ^{\ast }]  \notag \\
&=&\int \left\{ \frac{ib}{2}\left( \psi ^{\ast }\partial _{0}\psi -\partial
_{0}\psi ^{\ast }\cdot \psi \right) -\frac{e}{c}A_{0}-\frac{1}{2m}\left( -ib%
\mathbf{\nabla }\psi ^{\ast }-\frac{e}{c}\mathbf{A}\psi ^{\ast }\right)
\left( ib\mathbf{\nabla }\psi -\frac{e}{c}\mathbf{A}\psi \right) \right. 
\notag \\
&&\left. +\frac{b^{2}-\hbar ^{2}}{8\rho m}(\mathbf{\nabla }\rho )^{2}+\frac{%
b^{2}}{8m}\sum\limits_{\alpha =1}^{\alpha =3}(\mathbf{\nabla }s_{\alpha
})^{2}\rho \right\} \mathrm{d}^{4}x,  \label{s5.32}
\end{eqnarray}
where $\mathbf{s}$ and $\rho $ are defined by the relations (\ref{s5.30a}).
One should expect, that the two-component wave function describes the
general case, because the number of real components of the two-component
wave function coincides with the number of hydrodynamic variables $\left\{
\rho ,\mathbf{j}\right\} $. But this statement is not yet proved.

In the case of irrotational flow, when the two-component function $\psi $
has linear dependent components, for instance $\psi =\left\{ \psi
_{1},0\right\} $, the 3-vector $\mathbf{s}=$const, and the term containing
3-vector $\mathbf{s}$ vanishes. In the special case, when the
electromagnetic potentials $A_{k}=0$, the action (\ref{s5.32}) for $\mathcal{%
E}\left[ \mathcal{S}_{\mathrm{st}}\right] $ coincides with the action (\ref%
{a1.10}) for $\mathcal{S}_{\mathrm{S}}$.

Finally, if we choose the arbitrary constant $b$ in the form $b=\hbar $ and
set $A_{k}=0$, we obtain the action (\ref{a1.2}) for the Schr\"{o}dinger
particle.

\section{Addition after an attempt of this paper \\publication}

This paper has been submitted for publication to a scientific journal and
was rejected on the basis of the referee's report. The author disagrees with
the referee's remarks. He presents his comments to the referee's report in
the form of a dialogue, which is very effective for ventilation of the
truth. Unfortunately, according to regulations of Archives the dialogue form
of the paper is inadmissible for publication in archives. So, my comments
can be found on my personal web site (http://rsfq1.physics.sunysb.edu/%
\symbol{126}rylov/comme.htm). In general, the correspondence with the
scientific journal is confidential. In the given case the confidential
character of the correspondence is removed, because it is a correspondence
with an anonymous referee of an anonymous journal. Thus, the correspondence
contains only scientific component, which cannot be confidential.

\end{document}